\documentclass[pdflatex,sn-mathphys]{sn-jnl}
\jyear{2023}

\usepackage{rotating}
\usepackage{lineno}

\usepackage{caption}
\usepackage{multibib}
\newcites{supp}{Methods References}

\usepackage{afterpage}
\usepackage{comment}

\newcommand{\oiii}{[O\,{\sc iii}]}
\newcommand{\oii}{[O\,{\sc ii}]}
\newcommand{\sii}{[S\,{\sc ii}]}
\newcommand{\neiii}{[Ne\,{\sc iii}]}
\newcommand{\ciii}{C\,{\sc iii}]}
\newcommand{\nii}{[N\,{\sc ii}]}
\newcommand{\mgii}{Mg\,{\sc ii}]}

\begin{document}

\title[Spectroscopy of bright, early galaxy candidates]{Confirmation and refutation of very luminous galaxies in the early universe}



\author*[1]{\fnm{Pablo} \sur{Arrabal Haro}}
\author[1]{\fnm{Mark} \sur{Dickinson}}
\author[2]{\fnm{Steven L.} \sur{Finkelstein}}
\author[3]{\fnm{Jeyhan S.} \sur{Kartaltepe}}
\author[4]{\fnm{Callum T.} \sur{Donnan}}

\author[5]{\fnm{Denis} \sur{Burgarella}}
\author[4]{\fnm{Adam C.} \sur{Carnall}}
\author[4]{\fnm{Fergus} \sur{Cullen}}
\author[4]{\fnm{James S.} \sur{Dunlop}}
\author[6]{\fnm{Vital} \sur{Fern\'{a}ndez}}
\author[2,7,8]{\fnm{Seiji} \sur{Fujimoto}}
\author[9]{\fnm{Intae} \sur{Jung}}
\author[10]{\fnm{Melanie} \sur{Krips}}
\author[2,3,11]{\fnm{Rebecca L.} \sur{Larson}}
\author[12,13]{\fnm{Casey} \sur{Papovich}}
\author[14]{\fnm{Pablo G.} \sur{P\'{e}rez-Gonz\'{a}lez}}

\author[6,15]{\fnm{Ricardo O.} \sur{Amor\'{i}n}}
\author[2]{\fnm{Micaela B.} \sur{Bagley}}
\author[5]{\fnm{V\'{e}ronique} \sur{Buat}}
\author[2]{\fnm{Caitlin M.} \sur{Casey}}
\author[2,11]{\fnm{Katherine} \sur{Chworowsky}}
\author[16]{\fnm{Seth H.} \sur{Cohen}}
\author[9]{\fnm{Henry C.} \sur{Ferguson}}
\author[17]{\fnm{Mauro} \sur{Giavalisco}}
\author[18,19,20]{\fnm{Marc} \sur{Huertas-Company}}
\author[21,22]{\fnm{Taylor A.} \sur{Hutchison}}
\author[23]{\fnm{Dale D.} \sur{Kocevski}}
\author[9]{\fnm{Anton M.} \sur{Koekemoer}}
\author[9]{\fnm{Ray A.} \sur{Lucas}}
\author[4]{\fnm{Derek J.} \sur{McLeod}}
\author[4]{\fnm{Ross J.} \sur{McLure}}
\author[9]{\fnm{Norbert} \sur{Pirzkal}}
\author[5]{\fnm{Lise-Marie} \sur{Seill\'e}}
\author[24]{\fnm{Jonathan R.} \sur{Trump}}
\author[25]{\fnm{Benjamin J.} \sur{Weiner}}
\author[26,27]{\fnm{Stephen M.} \sur{Wilkins}}
\author[28]{\fnm{Jorge A.} \sur{Zavala}}

\affil[1]{NSF's National Optical-Infrared Astronomy Research Laboratory, 950 N. Cherry Ave., Tucson, AZ 85719, USA}
\affil[2]{Department of Astronomy, The University of Texas at Austin, Austin, TX, USA}
\affil[3]{Laboratory for Multiwavelength Astrophysics, School of Physics and Astronomy, Rochester Institute of Technology, 84 Lomb Memorial Drive, Rochester, NY 14623, USA}
\affil[4]{Institute for Astronomy, University of Edinburgh, Royal Observatory, Edinburgh EH9 3HJ, UK}
\affil[5]{Aix Marseille Univ, CNRS, CNES, LAM Marseille, France}
\affil[6]{Instituto de Investigaci\'{o}n Multidisciplinar en Ciencia y Tecnolog\'{i}a, Universidad de La Serena, Raul Bitr\'{a}n 1305, La Serena 2204000, Chile}
\affil[7]{Cosmic Dawn Center (DAWN), Jagtvej 128, DK2200 Copenhagen N, Denmark}
\affil[8]{Niels Bohr Institute, University of Copenhagen, Lyngbyvej 2, DK2100 Copenhagen \O, Denmark}
\affil[9]{Space Telescope Science Institute, Baltimore, MD, 21218, USA}
\affil[10]{IRAM, Domaine Universitaire, 300 rue de la Piscine, 38406 Saint-Martin-d'H\`{e}res, France}
\affil[11]{NSF Graduate Fellow}
\affil[12]{Department of Physics and Astronomy, Texas A\&M University, College Station, TX, 77843-4242 USA}
\affil[13]{George P.\ and Cynthia Woods Mitchell Institute for Fundamental Physics and Astronomy, Texas A\&M University, College Station, TX, 77843-4242 USA}
\affil[14]{Centro de Astrobiolog\'{\i}a (CAB), CSIC-INTA, Ctra. de Ajalvir km 4, Torrej\'on de Ardoz, E-28850, Madrid, Spain}
\affil[15]{Departamento de Astronom\'{i}a, Universidad de La Serena, Av. Juan Cisternas 1200 Norte, La Serena 1720236, Chile}
\affil[16]{School of Earth and Space Exploration, Arizona State University, Tempe, AZ, 85287 USA}
\affil[17]{University of Massachusetts Amherst, 710 North Pleasant Street, Amherst, MA 01003-9305, USA}
\affil[18]{Instituto de Astrof\'isica de Canarias, La Laguna, Tenerife, Spain}
\affil[19]{Universidad de la Laguna, La Laguna, Tenerife, Spain}
\affil[20]{Universit\'e Paris-Cit\'e, LERMA - Observatoire de Paris, PSL, Paris, France}
\affil[21]{Astrophysics Science Division, NASA Goddard Space Flight Center, 8800 Greenbelt Rd, Greenbelt, MD 20771, USA}
\affil[22]{NASA Postdoctoral Fellow}
\affil[23]{Department of Physics and Astronomy, Colby College, Waterville, ME 04901, USA}
\affil[24]{Department of Physics, 196 Auditorium Road, Unit 3046, University of Connecticut, Storrs, CT 06269, USA}
\affil[25]{MMT/Steward Observatory, University of Arizona, 933 N. Cherry Ave., Tucson, AZ 85721, USA}
\affil[26]{Astronomy Centre, University of Sussex, Falmer, Brighton BN1 9QH, UK}
\affil[27]{Institute of Space Sciences and Astronomy, University of Malta, Msida MSD 2080, Malta}
\affil[28]{National Astronomical Observatory of Japan, 2-21-1 Osawa, Mitaka, Tokyo 181-8588, Japan}


\maketitle 

\textbf{
During the first 500 million years of cosmic history, the first stars and galaxies formed, seeding the Universe with heavy elements and eventually reionizing the intergalactic medium \citep[][]{schneider02,wise12,jaacks19}.  Observations with \textbf{\textit{JWST}} have uncovered a surprisingly high abundance of candidates for early star-forming galaxies, with distances (redshifts, \boldmath{$z$}), estimated from multi-band photometry, as large as \boldmath{$z \approx 16$}, far beyond pre-\textbf{\textit{JWST}} limits \citep[][]{Castellano2022,Finkelstein2022a,Donnan2023,Harikane2023a,Atek2023,Finkelstein2023}.
While generally robust, such photometric redshifts can suffer from degeneracies and occasionally catastrophic errors.  Spectroscopic measurement is required to validate these sources and to reliably quantify physical properties that can constrain galaxy formation models and cosmology \citep{boylankolchin23}. Here we present \textbf{\textit{JWST}} spectroscopy that confirms redshifts for two very luminous galaxies with \boldmath{$z > 11$}, but also demonstrates that another candidate with suggested \boldmath{$z\approx 16$} instead has \boldmath{$z = 4.9$}, with an unusual combination of nebular line emission and dust reddening that mimics the colors expected for much more distant objects. These results reinforce evidence for the early, rapid formation of remarkably luminous galaxies, while also highlighting the necessity of spectroscopic verification. The large abundance of bright, early galaxies may indicate shortcomings in current galaxy formation models, or deviation from physical properties (such as the stellar initial mass function) that are generally believed to hold at later times.
}



\begin{figure}
    \centering
    \includegraphics[width=\linewidth]{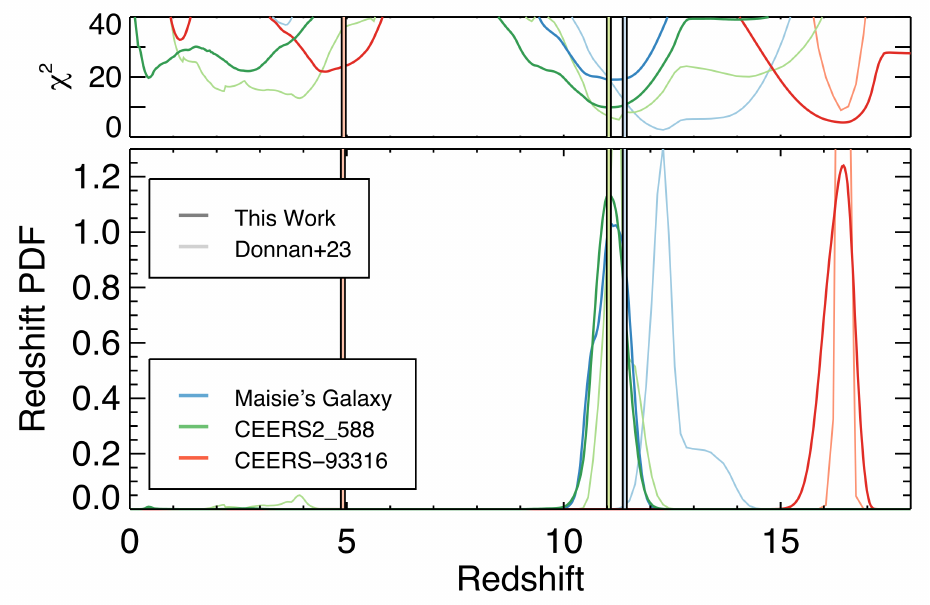}
    \caption{\textbf{Photometric redshift probability density function for the primary high-$\boldsymbol{z}$ candidates.} The likelihood (bottom) and $\chi^2$ goodness-of-fit (top) distribution functions for the photometric redshift, derived from the {\it JWST} CEERS photometry. Results for Maisie's Galaxy, CEERS2\_588 and the $z \sim$ 16.5 candidate galaxy CEERS-93316 are shown in blue, green and red, respectively. The line shading denotes the results using the photometric measurements from each team.  
    The spectroscopic measurements of $z = 11.416$, $z = 11.043$ and $z = 4.912$ from this paper are indicated by the thick vertical lines.}
    \label{fig:pz}
\end{figure}

The spectroscopic targets were selected from \textit{JWST}/NIRCam \citep{Rieke2023} imaging obtained through the Cosmic Evolution Early Release Science (CEERS) Survey \citep{bagley23} because their colors at wavelengths of $1 - 5$ microns matched predictions for galaxies with redshifts $z > 9$ \citep{Finkelstein2023}, including the key signature of a ``break'' in flux density at wavelengths shorter than that of redshifted hydrogen Lyman~$\alpha$ (Ly$\alpha$; rest-frame wavelength $\lambda_{\mathrm{rest}} = 1216$~\AA) due to the largely-neutral intergalactic medium (IGM) expected at early times. Here we focus attention on three bright extreme-redshift candidates: two with photometric redshifts $z \simeq 11$ (CEERS2\_5429, aka Maisie's Galaxy \citep{Finkelstein2022a, Finkelstein2023}, here MSA ID 1; and CEERS2\_588 \citep{Finkelstein2023}, here MSA ID 10), 
and another with a photometric redshift $z \simeq 16.4$ (CEERS-93316 \citep{Donnan2023}; here MSA ID 0). At their estimated redshifts, these three galaxies would have UV luminosities $\simeq 2$ to $\simeq 9$ times brighter than any other spectroscopically confirmed galaxy at $z > 11$ \citep{Robertson2023, curtis-lake2023}). We show these photometric redshift estimates in Figure~\ref{fig:pz}.

Although independent analyses by different teams \citep{Finkelstein2022a, Finkelstein2023, Donnan2023, Bouwens2023, Harikane2023a} identified these candidates using 7-band {\it JWST}/NIRCam photometry, it is possible for foreground objects to infiltrate high-redshift samples due to unusual colors resulting from older stellar populations, strong nebular emission lines, and/or dust reddening \citep{Donnan2023, Naidu2022b,perezgonzalez23,zavala23}.  A tentative detection at $850\,\mu$m of dust emission in the vicinity of CEERS-93316 \citep{zavala23} could support the dust reddening hypothesis, but we have investigated this using new observations at 1.1~mm with the NOEMA interferometer and localized the dust emission to another galaxy approximately 1.5~arcsec away (see Methods for details). Nonetheless, reddening could still affect the observed colors of CEERS-93316.

We observed these galaxies with {\it JWST}/NIRSpec \citep{Jakobsen2022} using its multi-object spectroscopy mode and prism disperser, whose low spectral resolution at wavelengths shorter than 2.5$\,\mu$m affords high sensitivity for detecting faint continuum emission and spectral breaks.  Details of the observations and data analysis, as well as results from simultaneous observations of several other candidate $z > 9$ galaxies,  are summarized in the Methods section. 

\begin{figure}
    \centering
    \includegraphics[width=\linewidth]{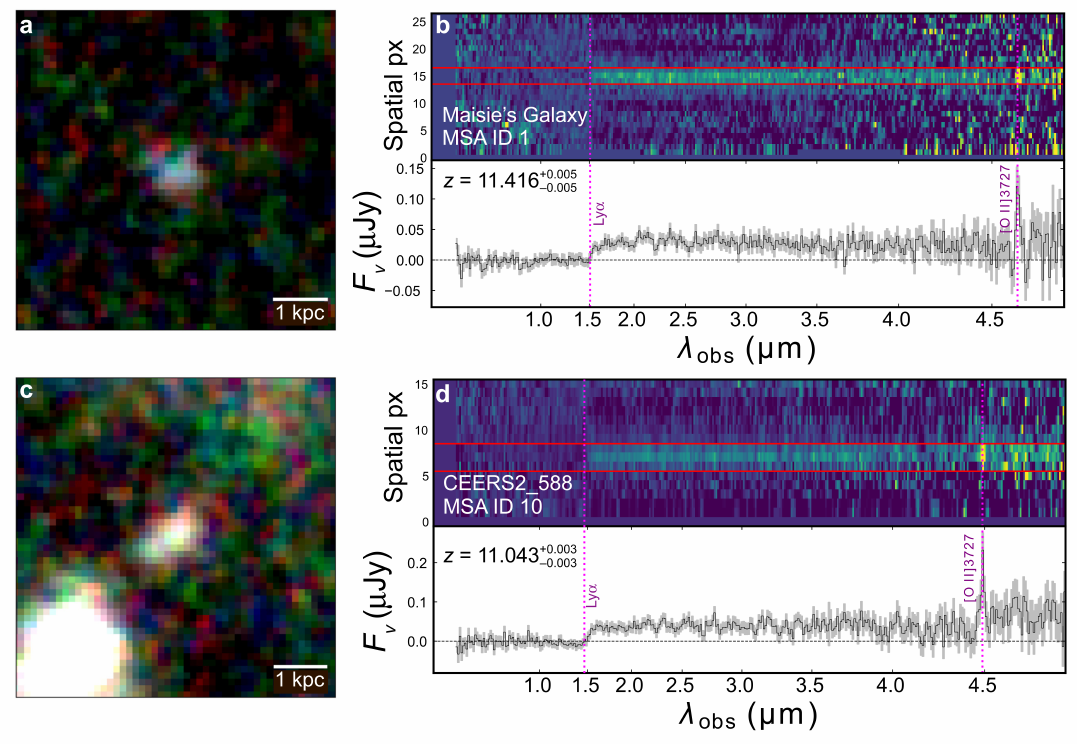}
    \caption{\textbf{Color stamps and spectra of the confirmed \boldmath{$z>11$} galaxies}. Panels a, c) Color images of Maisie's Galaxy (MSA ID 1) and CEERS2\_588 (MSA ID 10), showing approximately the true rest-frame ultraviolet color (made from the F277W, F356W, and F444W images).  
    Panels b, d) Spectra of Maisie's Galaxy and CEERS2\_588 highlighting the detected Ly$\alpha$ breaks and \oii\ emission lines.  The top panel shows the 2D spectrum.
    The red horizontal lines show the spatial extent over which the 1D spectrum, shown in the bottom panel, was extracted. The shaded grey area corresponds to 1$\sigma$ uncertainties in the 1D spectrum and the zero flux level is marked by a dashed black horizontal line. A clear spectral break is seen, below which there is no significant flux.  We interpret this as the Ly$\alpha$ break at a redshift of 11.416 and 11.043, which corresponds to a time 392 and 410 Myr after the Big Bang, respectively.}
    \label{fig:maisiebreak}
\end{figure}

The spectrum of Maisie's Galaxy (Figure~\ref{fig:maisiebreak}) shows well-detected continuum emission with a sharp break, below which no significant flux is detected.  The most likely interpretations of this feature are the Ly$\alpha$ IGM break or the Balmer break ($\lambda_{\mathrm{rest}}$ = 3646~\AA); the latter is prominent in galaxies where star formation ceased more than 100~Myr before the time of observation. Integrating the spectrum in wavelength intervals redder and bluer than the observed break, we measure a $3\sigma$ limit on the flux density ratio $> 2.6$, inconsistent with values expected for the Balmer break \citep{curtis-lake2023} (see Methods).
We used several methods \citep{ArrabalHaro2023} to measure the wavelength of the spectral break (see Methods and Extended Data Table~\hyperref[table:redshifts]{1}).  Interpreting this feature as the Ly$\alpha$ break (Figure~\ref{fig:maisiebreak}b) yields a redshift $ z = 11.44 ^{+0.09}_{-0.08}$, consistent with photometric redshift estimates. 
We also identify an emission line at 4.629$\,\mu$m as the unresolved singly-ionized \oii\ doublet ($\lambda_{\mathrm{rest}}$ = 3727,3730~\AA) at $z = 11.416$.

CEERS2\_588 exhibits a comparably strong Ly$\alpha$ break and \oii\ emission line at $z = 11.043$ \citep{Harikane2023b}. Its break amplitude is $>2.2$, again inconsistent with a Balmer break.  Another, fainter galaxy (MSA ID 64) also has a strong $(> 1.9)$ break and no emission lines, consistent with a Ly$\alpha$ break redshift $z = 10.10^{+0.13}_{-0.26}$ (see Methods).

NIRSpec observations of star-forming galaxies at $z < 9.6$ frequently detect strong \oiii\ ($\lambda_{\mathrm{rest}}$ = 4960,5008~\AA) and H$\beta$ ($\lambda_{\mathrm{rest}}$ = 4863~\AA) emission lines \citep{Fujimoto2023, RobertsBorsani2023, Williams2023} that shift beyond the instrument's spectral coverage for redshifts $z > 9.6$.  Emission lines at blue optical and ultraviolet rest-frame wavelengths are typically weaker and harder to detect in low signal-to-noise ratio (S/N) spectra. 
The \oii\ lines for Maisie's Galaxy and CEERS2\_588 are only weakly detected (4.9$\sigma$ and 6.4$\sigma$, respectively) even with moderately long NIRSpec exposure times, highlighting the importance of a robust Ly$\alpha$ break detection.

Maisie's Galaxy and CEERS2\_588 are the most UV-luminous galaxies with confirmed spectroscopic redshifts $z > 11$ to date, with rest-frame UV absolute magnitudes $M_{\mathrm{UV}} = -20.1$ and -20.3, respectively. 
The physical properties derived through stellar population modeling (see Methods for details) on the observed photometry and spectroscopy of Maisie's Galaxy and CEERS2\_588, fixing the redshifts to the spectroscopic values, are presented in Table~\ref{table:summary}. We find that they have 
stellar masses $\log_{10}(M_{\star}/M_{\odot}) = 8.6$ to 8.7 and low dust extinction ($A_V = 0.1$ to 0.2 mag), with high star-formation rates (SFR) for their stellar masses ($\log_{10}(\mathrm{sSFR}\equiv\mathrm{SFR}/M_{\star}) = -8.3$ to -7.7). 
These physical properties are consistent with similar values found for the few other spectroscopically confirmed galaxies at $z\gtrsim11$ \citep{Robertson2023, Tacchella2023}.

\begin{table}
\begin{center}{\bf Table 1: Summary of galaxy properties}\end{center}
\begin{center}
\begin{tabular}{lccc}
\hline
\hline
Property & MSA ID 1 & MSA ID 10 & MSA ID 0\\
  & (Maisie's Galaxy)  & (CEERS2\_588) & (CEERS-93316) \\
\hline
RA (J2000) & 214.943152 & 214.906640 & 214.914550 \\
Dec (J2000) & 52.942442 & 52.945504 & 52.943023 \\
Photometric Redshift & 11.08$^{+0.36}_{-0.39}$ & 11.02$^{+0.39}_{-0.27}$ & 16.45$^{+0.45}_{-0.18}$ \\
Spectroscopic Redshift & 11.416$\pm$0.005 & 11.043$\pm$0.003 & 4.912$\pm$0.001 \\
$\mathcal{T}_{\mathrm{BigBang}}$ (Myr) & 392 & 410 & 1194 \\
$M_{\mathrm{UV}}$ & $-20.1^{+0.1}_{-0.1}$ & $-20.3^{+0.1}_{-0.2}$ & $-16.2^{+0.5}_{-0.1}$ \\
\hline
$\log(M_{\star}/M_{\odot}$) & $8.6 \pm 0.3$ & $8.7 \pm 0.1$ & $8.8 \pm 0.1$ \\
\hspace{0.3cm}Range: & 8.0 -- 9.0 & 8.1 -- 9.1 & 8.3 -- 9.2 \\
$A_V$ (mag) & $0.1 \pm 0.1$ & $0.2 \pm 0.1$ & $2.3 \pm 0.2$ \\
\hspace{0.3cm}Range: & 0.0 -- 0.8 & 0.03 -- 0.7 & 0.8 -- 2.7 \\
SFR ($M_\odot$~yr$^{-1}$) & $2 \pm 1$ & $10 \pm 4$ & $60 \pm 20$ \\
\hspace{0.3cm}Range: & 1 -- 12 & 2 -- 17 & 2 -- 80 \\
$\log(\mathrm{sSFR/yr^{-1}})$ & $-8.3^{+0.2}_{-0.3}$ & $-7.7^{+0.2}_{-0.2}$ & $-7.0^{+0.1}_{-0.2}$ \\
\hspace{0.3cm}Range: &  -8.6 -- -7.6   &  -8.2 -- -7.5    &  -8.1 -- -6.9 \\
\hline
\end{tabular}
\caption{The photometric redshift calculation is described in the Methods and has been updated for this work. Spectroscopic redshifts are derived from emission lines; see Methods for a full discussion.
$\mathcal{T}_{\mathrm{BigBang}}$ denotes the age of the universe at the spectroscopic redshift.  
All uncertainties are 68\% confidence intervals.
Stellar masses ($M_\star$), SFRs averaged over the last 100 Myr and $A_V$ values have been derived using three different stellar population modeling techniques (see Methods). For each property, the first line gives the best-fit values and 68\% confidence intervals from the method simultaneously fitting both the spectra and the photometry (Synthesizer), while the second line shows the ranges spanned by all three modeling methods including their 68\% confidence intervals.}\label{table:summary}
\end{center}
\end{table}

\begin{figure}
    \centering
    \includegraphics[width=\linewidth]{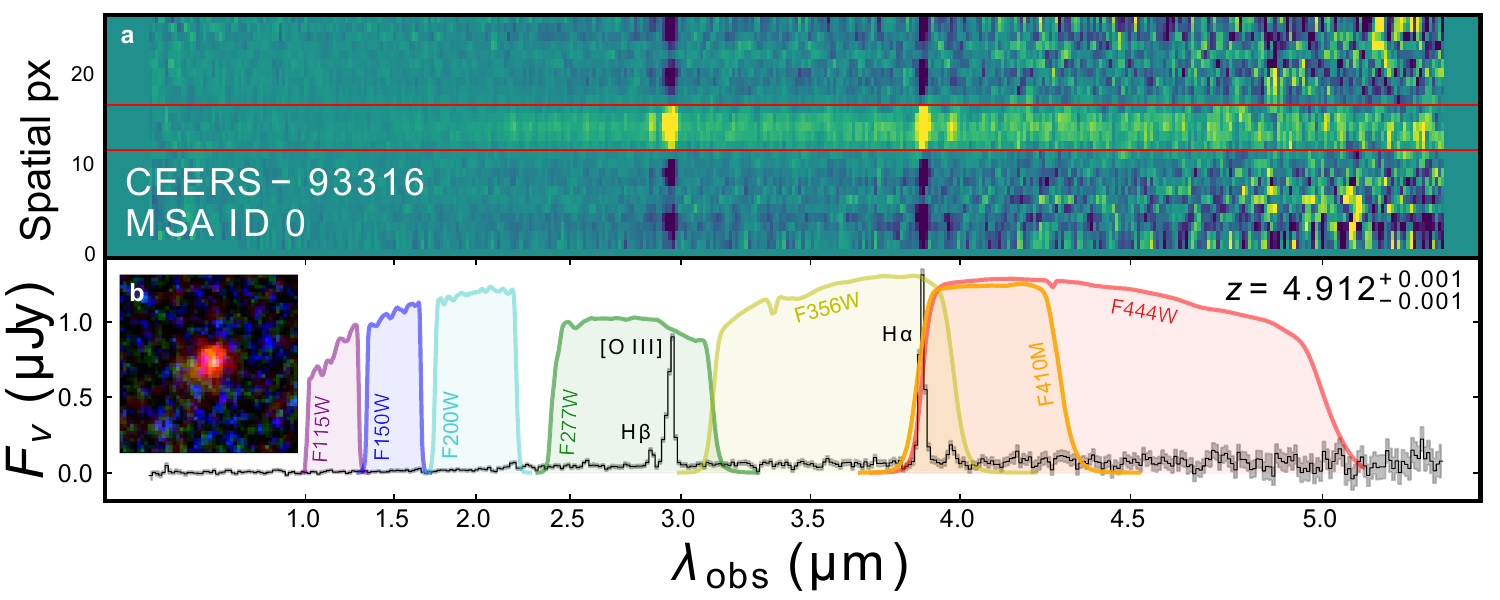}
    \caption{\textbf{NIRSpec spectrum of \boldmath{$z_{\mathrm{phot}}\sim16$} candidate CEERS-93316 (MSA ID 0).}  
    Panel a) shows the 2D spectrum with the red lines denoting the spatial extraction limits, and panel c) shows the 1D extraction. Emission lines labeled in panel c) indicate an unambiguous $z=4.912$ for this object. The shaded grey area corresponds to 1$\sigma$ uncertainties in the 1D spectrum. Panel c) also illustrates the wavelengths covered by the {\it JWST}/NIRCam filter set used to select this galaxy, highlighting how \oiii\ contamination in F277W (green) and H$\alpha$ contamination in \emph{all} of F356W (yellow), F410M (orange) and F444W (red) conspired to mimic the signature of both a very high-redshift Ly$\alpha$ break and a blue continuum at longer wavelengths.  This unusual (but educational) situation is detailed further in  Figure~\ref{fig:sed}.  Panel b) shows a F150W$+$F200W$+$F277W color image of this galaxy.}
    \label{fig:donnanspectrum}
\end{figure}

\begin{figure}
    \centering
    \includegraphics[width=\linewidth]{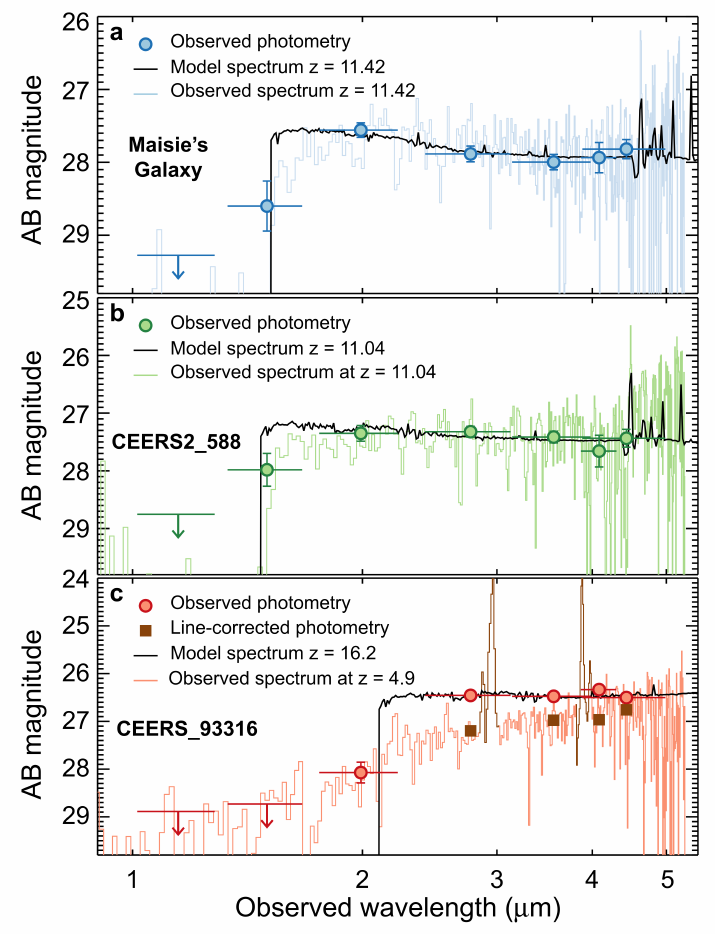}
    \caption{\textbf{Spectral energy distributions of the three main sources reported.} For a) Maisie's Galaxy (MSA ID 1), b) CEERS2\_588 (MSA ID 10) and c) CEERS-93316 (MSA ID 0), each panel shows the observed photometry (colored points) compared to the best-fitting stellar population model (black line) and the observed spectra (colored line).  In panel c), the model is deliberately shown at $z =$ 16.2 to demonstrate the way in which strong emission lines impact the observed photometry in this specific case to replicate the presence of the Ly$\alpha$ break at $z \simeq 16$ and an apparently blue continuum at longer wavelengths.  The brown squares show the observed photometry omitting the spectroscopic emission line contribution (shaded brown in the spectrum). Error bars are 1$\sigma$ uncertainties while the upper limits correspond to 2$\sigma$ limits.}
    \label{fig:sed}
\end{figure}

The spectrum of CEERS-93316 (Figure \ref{fig:donnanspectrum}) shows emission lines from the hydrogen Balmer series, \oiii, \oii, and \sii\ ($\lambda_{\mathrm{rest}}$ = 6718,6733~\AA) at a redshift $z = 4.912 \pm 0.001$.  This is the same redshift as for another galaxy, CEERS-DSFG-1 (MSA ID 2) \citep{Barrufet2023, zavala23}, which lies 26 arcsec away and was previously discussed as a dusty, star forming galaxy with colors that could mimic a high-redshift Ly$\alpha$ break \citep{zavala23}. 
At least two other neighboring galaxies have very similar redshifts (see Methods).
The continuum emission of CEERS-93316 fades at wavelengths shorter than 2.5$\,\mu$m (rest frame $\lesssim 0.4\,\mu$m), although flux is significantly detected in the spectrum down to $\lambda \approx 1.7\,\mu$m, and marginally detected down to $\lambda \approx 1.3\,\mu$m. The strong H$\alpha$+[NII] and \oiii+H$\beta$ lines have large rest-frame equivalent widths (EWs; $760 \pm 67$~\AA\ and $1116 \pm 13$~\AA, respectively), and hence contribute significantly to the flux measured by NIRCam through the F277W, F356W, F410M, and F444W filters. At this precise redshift, H$\alpha$ emission actually contributes to the flux in the last three of those filters (Figure~\ref{fig:donnanspectrum}). The combination of the strong lines and the red continuum leads to the red colors measured between filters F150W, F200W, and F277W and the blue colors from F277W to F444W (Figure~\ref{fig:sed}c).

The significant reddening implied by the faint, red UV continuuum emission is corroborated by the large Balmer decrement measured from the spectrum (H$\alpha$/H$\beta=5.2\pm0.5$), which implies -- despite the large EWs of the lines -- a substantial nebular color excess of $\mathrm{E(B-V)}=0.60\pm0.10$ (see Methods).
Indeed this combination of line emission and reddening at $z \simeq 5$ has been considered previously as an alternative explanation for the colors of CEERS-93316 \citep{Donnan2023, zavala23, Naidu2022b, perezgonzalez23}.
The H$\alpha$ emission may imply rapid star formation ($28^{+7}_{-6}\ M_\odot$~yr$^{-1}$, see Methods), although some other faint red objects at $z \sim 5$ show strong line emission with broad Balmer line components that indicate the presence of active galactic nucleii \citep[e.g.,][]{kocevski23, Matthee2023}.

The example of CEERS-93316 is a reminder that the prevalence of very strong nebular line emission in early galaxies can influence broad-band photometry with instruments such as {\it JWST}/NIRCam and mimic the colors predicted for much more distant objects. Exceptional candidates for galaxies at extreme redshifts require spectroscopic confirmation. However, we note that this is a fairly extreme case, as it is only at very specific redshifts that strong emission lines can fully mimic a Ly$\alpha$ break with multiple broad-band filter detections, and reddening is also required to suppress detection in bluer filters (see Methods).

Fortunately {\it JWST}/NIRSpec is a powerful tool for spectroscopic analysis and redshift measurement, as demonstrated by our verification of two galaxies with $z > 11$ and three other galaxies at $7.9 < z < 10.1$ (discussed in Methods).  NIRSpec observations have now confirmed redshifts for 10 galaxies at $z > 10$ \citep{ArrabalHaro2023, bunker2023, curtis-lake2023, Harikane2023b, Hsiao2023}, including five with $M_\mathrm{UV} < -20$, unexpectedly bright compared to most pre-{\it JWST} predictions. These confirmations are crucial to validate the methods employed for $z\gtrsim10$ photometric selection \citep{ArrabalHaro2023} and to improve our physical understanding of the first phases of galaxy formation.  

Early analyses of \emph{candidate} galaxies at $z = 9.5$--12 have shown that their observed abundance is significantly higher than predicted by nearly all theoretical models \citep{Finkelstein2022a,Finkelstein2023,Harikane2023a,Leung2023}.  This could indicate that specific assumptions in these models need to be modified at very high redshifts, including the possibility of negligible dust attenuation \citep{Ferrara2023, Mason2023} or an increased efficiency of converting gas into stars, or even a shift toward the formation of more massive stars in these early epochs.  
Such a ``top-heavy" initial mass function (IMF) has indeed been predicted to be present in early galaxies, driven by a combination of their chemically simple makeup \citep{bromm01,sharda22} as well as a rising cosmic microwave background temperature floor \citep{larson98}.  However, before any such unusual physical properties can be robustly inferred for early galaxies, their photometric redshift estimates must be verified.  
Four out of seven photometric candidates from CEERS data at $9.5 < z < 12$ with $M_{UV} < -20$ have been spectroscopically confirmed \cite[]{ArrabalHaro2023}, including the present examples at $z > 11$, with no confirmed interlopers except the $z \simeq 16$ candidate CEERS-93316. The space density of confirmed objects alone exceeds the predictions of most theoretical models.
The spectroscopic observations reported here reinforce the higher-than-expected abundance of such bright galaxies in the very early universe, while also vetting photometric redshift outliers like CEERS-93316 that could be misinterpreted if left unrecognized.


\newpage

\clearpage

\captionsetup[table]{name=Extended Data Table}
\captionsetup[figure]{name=Extended Data Figure}
\setcounter{figure}{0}    
\setcounter{table}{0}

\section*{Methods}\label{sec_methods}

\subsection*{Cosmological model}\label{sec_cosmo_params}

Throughout this work we adopt a $\Lambda$ cold dark matter cosmological model with parameters $\Omega_\mathrm{m} = 0.315$, $\Omega_\mathrm{\Lambda} = 0.685$ and $H_{0}=67.36$ km s$^{-1}$ Mpc$^{-1}$ \citesupp{Planck2020}.
All magnitudes are presented in the AB system \citesupp{Oke1983}.

\subsection*{Near-infrared imaging data and spectroscopic target selection}\label{sec_imagdata}

We select high redshift galaxy candidates from the first epoch of imaging from the Cosmic Evolution Early Release Science (CEERS; PI: S. Finkelstein \citep{Finkelstein2022a}) survey.  These data consist of four roughly contiguous pointings with the NIRCam instrument on {\it JWST}, covering an area of $\sim$35 arcmin$^2$.  The full area is covered by six broad-band filters (F115W, F150W, F200W, F277W, F356W, F444W) and one medium-band filter (F410M).  We use the v0.5 data release from the CEERS team\footnote{\url{https://ceers.github.io}} \citep{bagley23}.  These reductions use the STScI {\it JWST} pipeline with several custom modifications, correcting for various issues, including $1/f$ noise, wisps, and snowballs\footnote{\url{https://jwst-docs.stsci.edu/data-artifacts-and-features/snowballs-and-shower-artifacts}}, with updated photometric calibration from late 2022.  The 5$\sigma$ photometric depths of these data for compact source detection are $\sim$29-29.2 AB magnitudes in F115W, F150W, F200W, F277W and F356W, and $\sim$0.7 and $\sim$0.5 mag shallower in F410M and F444W, respectively.

Photometry was optimized for accurate measures of colors, generally following previous CEERS measurements \cite{Finkelstein2023}, with a few key updates to improve measured colors, particularly between {\it HST} and {\it JWST}.  Specifically, all bands with point spread functions (PSFs) smaller than that of F277W (i.e., ACS F606W, F814W, and NIRCam F115W, F150W, and F200W) were convolved to match the F277W PSF.  For the remaining images, fluxes were measured in the native images, but a correction was applied on a per-source basis as the ratio of the flux in the native F277W image to that of the F277W image convolved to match the PSF in a given larger-PSF image.   We then used simulation-based corrections to derive accurate total fluxes.  Photometric redshifts were derived using {\sc eazy} \citepsupp{brammer08}, including new custom templates designed to match the expected blue colors of early galaxies \citepsupp{larson22}. 

The NIRSpec spectroscopic observations targeted three primary galaxy candidates: CEERS2\_5429, aka ``Maisie's Galaxy'' \citep{Finkelstein2022a, Finkelstein2023}, with photometric redshift $z \approx 11.1$, ``CEERS-93316'' with photometric redshift $z \approx 16.4$ \citep{Donnan2023}, and ``CEERS-DSFG-1'', a {\it JWST}-detected galaxy with millimeter-wavelength emission that is likely to arise from dust-obscured star formation at $z \sim 5$ \citep{Barrufet2023}, but whose near-infrared colors could be interpreted as suggesting $z >$ 10 \citep{zavala23}.
Many other faint, high-redshift candidates were also observed simultaneously, including CEERS2\_588, which has photometric redshift $z \approx 11$ \citep{Donnan2023, Finkelstein2023} and is comparably bright as Maisie's Galaxy.
These object names come from the original papers that reported these objects rather than from a homogeneous source catalog.

\subsection*{NOEMA data and analysis}\label{sec_noemadata}

Observations with the JCMT and its SCUBA2 bolometer array detected tentative 850$\,\mu$m emission in the vicinity of CEERS-93316. 
Previous analysis \citepsupp{zavala23} showed that this submillimeter emission is unlikely to arise from an object at $z > 10$, and that a $z\sim 5$ galaxy with dust attenuation and strong emission lines could explain the measured {\it JWST} photometry.

We observed dust continuum with band 3 of the Institute Radio Astronomie Millim\'etrique (IRAM) NOrthern Extended Millimeter Array (NOEMA) as a Director's Discretionary Time (DDT) program \#D22AC001 (PI: S.\ Fujimoto). The observations were carried out between 2022 Nov 26 and 2022 Dec 08 in several visits with the C configuration using 10--11 antennas. The data were processed in the standard manner with the pipeline using the latest version of GILDAS. We used CASA version 6.5 for the imaging \citesupp{CASA2022}. 

The upper and lower side bands of the band 3 receiver were tuned to 256.6~GHz and 271.1~GHz, resulting in a central wavelength of 1.134~mm. 
The observations of 3C45, 3C273, J1439+499, and 3C84 served as band-pass phase calibrators. 
Additional targets 1418+546 and J1439+499 were used for the phase and amplitude calibrations. We calibrated the absolute flux scale against MWC349, LkHa101, and 2010+723 whose flux densities are regularly monitored at NOEMA. The total integration time on-source was 6.5 hours. 

To maximize sensitivity, we used natural weighting for the imaging. The resulting 1~mm continuum map has synthesized beam FWHM of $0.^{\prime\prime}91\times0.^{\prime\prime}67$ with $1\sigma$ sensitivities of $28\,\mu\mathrm{Jy}~\mathrm{beam}^{-1}$ for the continuum. We do not identify any line features at the target position and thus produce the 1~mm continuum map from all channels. 

\begin{figure}
    \centering
    \includegraphics[width=0.5\linewidth]{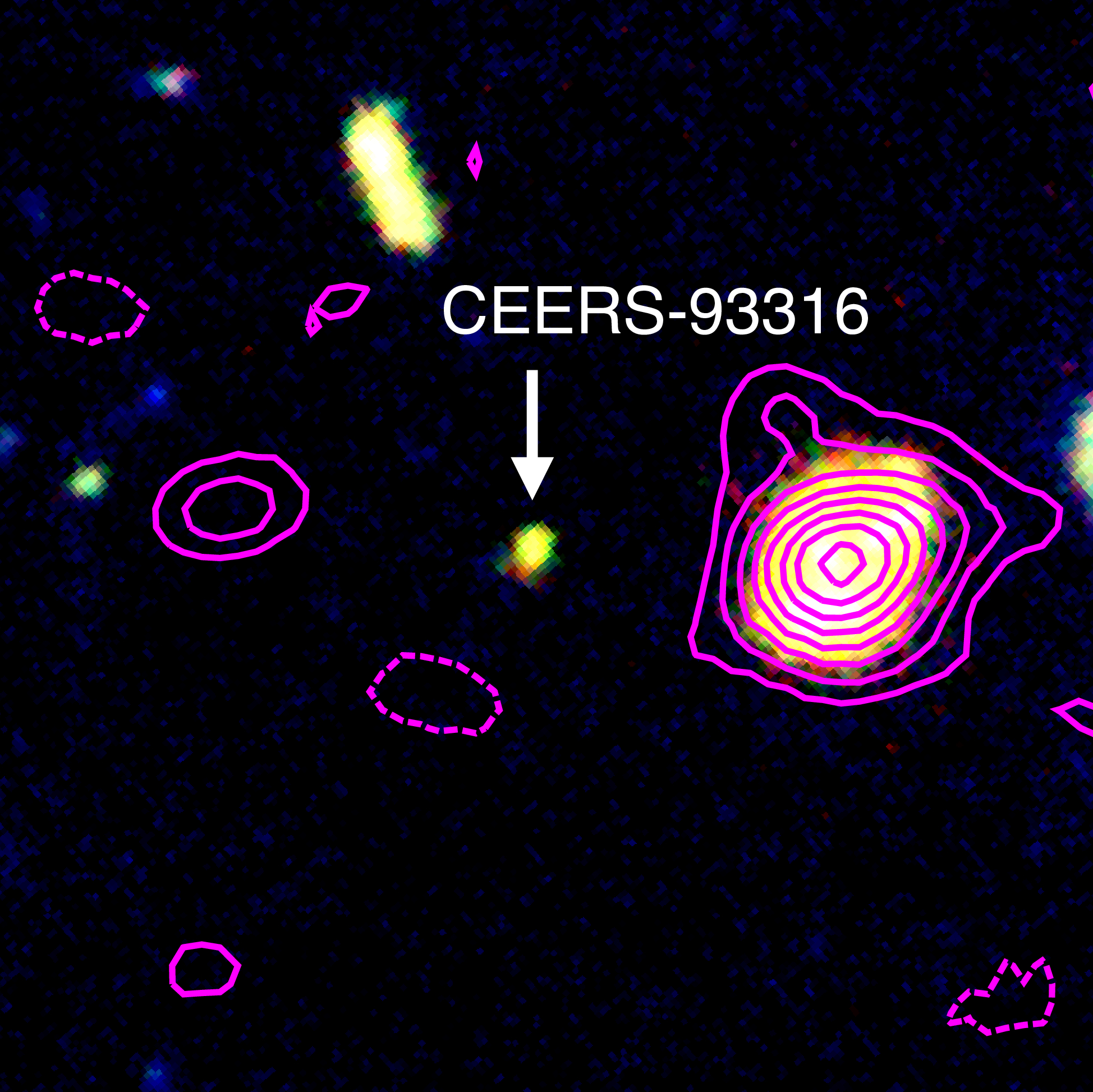}
    \caption{NOEMA 1.1-mm contours overlaid on the NIRCam $6''\times6''$ RGB color image (R: F444W, G: F356W, B: F200W) around CEERS-93316 (MSA ID 0) pointed by the white arrow. North is up and East is left. The magenta contours are drawn at 1$\sigma$ intervals from $\pm2\sigma$ to $\pm9\sigma$. 
    CEERS-93316 is undetected at 1.1-mm, while robust emission from a nearby source is likely responsible for a previously reported tentative SCUBA2 detection \cite{zavala23}).}
    \label{fig:noema}
\end{figure}

In Extended Data Figure~\ref{fig:noema}, we show the NOEMA 1.1-mm intensity contours overlaid on the NIRCam color image around CEERS-93316. We do not detect emission at the position of CEERS-93316, with a 3$\sigma$ upper limit of 0.084~$\mathrm{mJy}~\mathrm{beam}^{-1}$.

On the other hand, the NOEMA observations revealed relatively bright continuum emission at 9$\sigma$ significance from a nearby ($\sim1.5^{\prime\prime}$) galaxy with a photometric redshift $z \approx 4.9$, very similar to the spectroscopic redshifts of CEERS-93316, CEERS-DSFG-1 (also detected at millimeter wavelengths), and other several galaxies in the field (see below and Extended Data Table~\hyperref[table:redshifts]{1}). 
We measure a 1.1~mm flux density $0.56\pm0.07$~mJy with an optimized aperture. This is in good agreement with the previous report of SCUBA2 850~$\mu$m emission with $0.64\pm0.26$~mJy \citesupp{zavala23}, implying that this neighboring galaxy is the main contributor to the SCUBA2 detection.  

\subsection*{Near-infrared spectroscopy and data reduction}\label{sec_specdata}

{\it JWST}/NIRSpec multi-object spectroscopic observations (DDT program \#2750; PI: P.\ Arrabal Haro) were executed on UTC 2023 March 24-25 using the PRISM/CLEAR combination of disperser and filter.  The prism disperser covers the wavelength range 0.6 to $5.3\,\mu$m with varying spectral resolution $R \equiv \lambda/\Delta\lambda \approx 30$ at $\lambda = 1.2\,\mu$m to $>300$ at $\lambda > 5\,\mu$m.  The low resolution of the prism at bluer wavelengths aids the detection of faint UV continuum and the Ly$\alpha$ break, while the higher resolution at red wavelengths facilitates the detection of redshifted optical rest-frame emission lines.

The NIRSpec Micro-Shutter Assembly (MSA) \citepsupp[][]{Ferruit2022} was configured to observe 160 faint-object targets. 
MSA ID numbers were assigned to the targets when the spectroscopic observing plan was designed.
For most observed targets, three-shutter slitlets were employed, enabling a three-point nodding pattern to facilitate background subtraction. 
Exceptionally, CEERS2\_588 was fit into the MSA configuration ``by hand'' using a two-shutter slitlet, and therefore received two thirds of the full exposure time.  
The observations consisted of 3 sequences of 3 nodded integrations, each with 9 to 10 readout groups in NRSIRS2 readout mode, for a total combined exposure time of 18387~s.

The MSA configurations included a total of 7 NIRCam-selected $z\gtrsim8.5$ galaxy candidates as well as 153 other targets.  We report results for 10 objects in Extended Data Table~\hyperref[table:redshifts]{1} including the primary targets described above.
Essential parameters for Maisie's Galaxy (MSA ID 1), CEERS2\_588 (MSA ID 10), and CEERS-93316 (MSA ID 0), including results from the present analysis, are summarized in Table~\ref{table:summary}. Detailed analysis of CEERS-DSFG-1 (MSA ID 2) will be presented elsewhere.

\begin{table}
\setlength{\tabcolsep}{3pt}
\begin{center}{\bf  Extended Data Table 1: Summary of spectroscopically confirmed galaxies}\end{center}
\begin{center}
\begin{tabular}{lccccccc}
\hline
\hline
 MSA & R.A.\ & Dec & $\mathrm{F277W}$ & $z_{\mathrm{phot}}$ & $z^{\mathrm{lines}}_{\mathrm{spec}}$ &  $z^{\mathrm{break}}_{\mathrm{spec}}$ &  $z^{\mathrm{cont}}_{\mathrm{spec}}$ \\
 ID & (J2000) & (J2000) & (AB) & 68\% C.L. & &  & \\
\hline
1 & 214.943152 & 52.942442 & 27.9 & 10.7--11.5 & $11.416\pm0.005$ & 11.44$^{+0.09}_{-0.08}$ & 11.38$^{+0.14}_{-0.08}$\\
10 & 214.906640 & 52.945504 & 27.3 & 10.6--11.3 & $11.043\pm0.003$ & 11.43$^{+0.01}_{-0.07}$ & 11.03$^{+0.02}_{-0.02}$ \\
64 & 214.922787 & 52.911529 & 28.3 & 10.4--11.6 & --- & 10.10$^{+0.13}_{-0.26}$ & 9.73$^{+0.21}_{-0.13}$\\
28 & 214.938642 & 52.911749 & 26.9 & 8.90--9.00 & 8.763 $\pm$ 0.001 & 8.90$^{+0.04}_{-0.02}$ & 8.72$^{+0.01}_{-0.01}$\\
355 & 214.944766 & 52.931450 & 28.7 & 8.0--8.5 & 7.925 $\pm$ 0.001 & 7.71$^{+0.12}_{-0.10}$ & 7.88$^{+0.01}_{-0.01}$\\
0 & 214.914550 & 52.943023 & 26.5 & 16.0--16.6 & 4.912 $\pm$ 0.001 & --- & 4.89$^{+0.01}_{-0.01}$\\
2 & 214.909116 & 52.937205 & 26.2 & 3.4--13.6 & $4.910\pm0.003$ & --- & 4.88$^{+0.01}_{-0.01}$\\
2763 & 214.927789 & 52.935859 & 25.1 & 4.87--5.44 & 4.902 $\pm$ 0.001 & --- & 4.87$^{+0.01}_{-0.01}$\\
3149 & 214.914917 & 52.943622 & 24.7 & 0.88--4.81 & 4.901 $\pm$ 0.001 & --- & 4.87$^{+0.01}_{-0.01}$\\
69 & 214.861602 & 52.904604 & 28.2 & 9.5--10.3 & --- & --- & ---\\
\hline
\end{tabular}
\caption*{\textbf{Extended Data Table 1:} The photometric redshifts limits correspond to 68\% confidence intervals. Spectroscopic redshifts presented are based on fitting of emission lines, the Ly$\alpha$ break, and the complete  continuum measured with the NIRSpec prism.
Maisie's Galaxy, CEERS2\_588 and CEERS-93316 correspond to MSA IDs 1, 10 and 0, respectively.}\label{table:redshifts}
\end{center} 
\end{table}

The NIRSpec data processing followed the methodology employed for other CEERS NIRSpec prism observations \citep{ArrabalHaro2023, Fujimoto2023}.  We make use of the STScI Calibration Pipeline\footnote{\url{https://jwst-pipeline.readthedocs.io/en/latest/index.html}} version 1.8.5 and the Calibration Reference Data System (CRDS) mapping 1088. We use the \texttt{calwebb\_detector1} pipeline module to subtract the bias and the dark current, correct the 1/$f$ noise, and generate count-rate maps (CRMs) from the uncalibrated images. The parameters of the \texttt{jump} step are modified for an improved correction of ``snowball'' 
events associated with high-energy cosmic rays.

The resulting CRMs are then processed with the \texttt{calwebb\_spec2} pipeline module which creates two-dimensional (2D) cutouts of the slitlets, performs the background subtraction making use of the nodding pattern, corrects the flat-fields, implements the wavelength and photometric calibrations and resamples the 2D spectra to correct the distortion of the spectral trace. The \texttt{pathloss} step accounts for the slit loss correction at this stage of the reduction process.

The images of the individual nods are combined at the \texttt{calwebb\_spec3} pipeline stage, making use of customized apertures for the extraction of the one-dimensional (1D) spectrum. 

The 2D and 1D spectra are simultaneously inspected with the Mosviz visualization tool\footnote{\url{https://jdaviz.readthedocs.io/en/latest/mosviz/index.html}} \citepsupp{JDADF2023} to mask possible remaining hot pixels and other artifacts in the images, as well as the detector gap (when present).  Data from the three consecutive exposure sequences are combined after masking image artifacts to produce the final 2D and 1D spectral products.

The \textit{JWST} pipeline uses an instrumental noise model to calculate flux errors for the extracted spectra.  We test and rescale these flux errors for the effect of data correlation introduced by the pipeline following procedures described elsewhere \citep{ArrabalHaro2023}.
To further test possible flux discrepancies due to uncorrected slit losses or inaccurate flux calibration, we integrate the spectra through the NIRCam filter bandpasses and compare the resulting synthetic photometry to the measured NIRCam photometry. For the main sources presented in this work, the flux ratios are similar for all NIRCam bands with robust (5$\sigma$) detections, and do not show any trend with wavelength. We rescale the spectroscopic fluxes by a constant factor derived from the average of the flux ratios for all bands with detections. These rescaling factors are 1.29, 1.46 and 1.56 for CEERS-93316 (MSA ID 0), Maisie's Galaxy (MSA ID 1) and CEERS2\_588 (MSA ID 10), respectively.

\subsection*{Redshift estimation}\label{sec:redshift_estimation}

For objects with multiple clear emission lines, we were easily able to identify the \oiii\ and/or H$\alpha$ emission lines.  Spectroscopic redshifts are measured by comparing the observed emission lines with their theoretical vacuum transitions. A Gaussian profile is fitted for every spectral line to establish the transition wavelength. The Gaussian fluxes are used to weigh the mean redshifts in Extended Data Table~\hyperref[table:redshifts]{1}. In those observations, where the \oiii\ doublet lines are merged, the profile fitting assumes the same gas kinematics and their amplitudes fixed by the theoretical emissivity ratio (2.98 \citepsupp{Storey_2004_O3}) to isolate them. 

We also estimate the redshifts using the detected continuum with two methods. The first method consists of fitting a five-parameter model to describe an object's spectrum and determine the wavelength of the Ly$\alpha$ break \citesupp{ArrabalHaro2023}.
We first create the intrinsic spectrum based on the redshift, UV absolute magnitude, and UV spectral slope.  We then add IGM absorption by zeroing out the spectrum below 1215.67 \AA\ rest-frame wavelength, and add Ly$\alpha$ damping wing absorption by adding two additional parameters: the neutral hydrogen fraction ($x_{\mathrm{HI}}$) and an ionized bubble radius ($R_{\mathrm{bubble}}$) \citesupp{dijkstra14, curtis-lake2023}.  We derive posterior constraints on these five parameters using an IDL implementation of the \textsc{emcee} Python code \citesupp{finkelstein19}, restricting the spectrum to wavelengths below 2500 \AA\ rest-frame for a given redshift.  
The second method is full spectral fitting to the NIRSpec data, which detects the continuum for all our galaxies, using {\sc eazy} \citepsupp{brammer08} and a variety of spectral templates, including UV-bright and dusty galaxies. 
For Maisie's Galaxy and CEERS2\_588 the spectra cover both the Lyman and Balmer breaks. All methods provide results in agreement within $\Delta z\sim0.1$, consistent within our estimated uncertainties.

\subsection*{Redshift results}\label{sec:redshift_results}

\begin{figure}
    \centering
    \includegraphics[width=\linewidth]{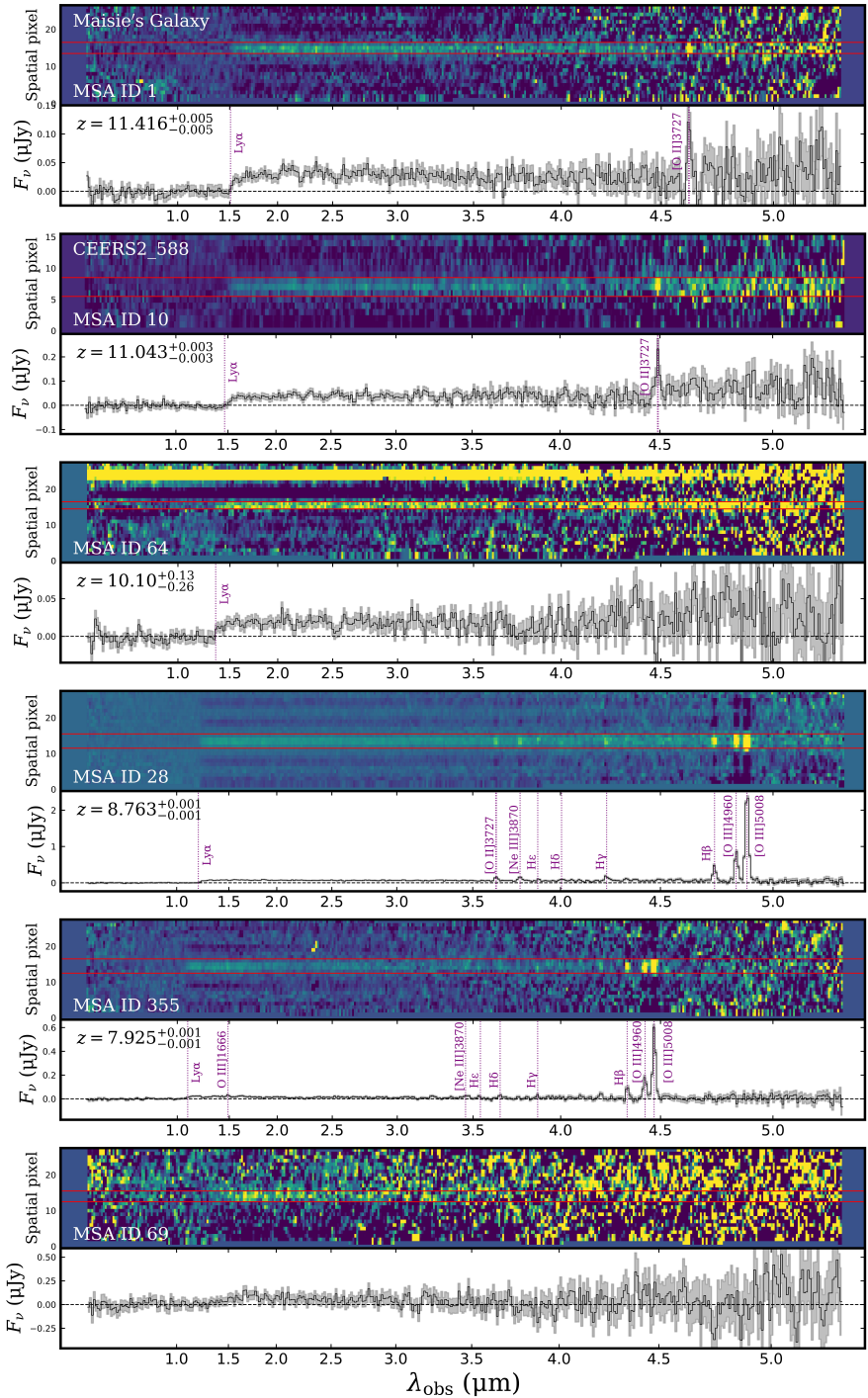}
    \caption{2D and 1D NIRSpec spectra of six high-$z$ candidates.  Features and spectroscopic redshifts are indicated, except for MSA ID 69 where we cannot determine a redshift with confidence. The shaded grey area corresponds to 1$\sigma$ uncertainties.}
    \label{fig:highz_spectra}
\end{figure}

We derive redshifts from multiple emission lines  for galaxies with MSA IDs 0 (via \oii, \oiii, H$\alpha$), 2 (via \oiii, H$\alpha$),  28 (via \oii, \oiii) and 355 (via \oiii,  H$\beta$).  We measure Ly$\alpha$ break redshifts for MSA IDs 1, 10, 28, 64 and 355. For MSA IDs 1 and 10 we also detect single emission lines that we identify as the unresolved \oii\ doublet at redshifts that are consistent with (but more precise than) the Ly$\alpha$ break  redshifts, and we adopt those values as our fiducial redshifts for those galaxies.
Source ID 69 shows a weak break at $\sim$1.3~$\mu$m, and this methodology does find a maximum likelihood of $z =$ 9.744.  However, the result has a large uncertainty, with a significant probability of the redshift being lower.  We thus do not consider MSA ID 69 confirmed. We present our redshift measurements in Extended Data Table~\hyperref[table:redshifts]{1} and the spectra of the confirmed high-$z$ objects are shown in Extended Data Figure~\ref{fig:highz_spectra}.

\begin{figure}
    \centering
    \includegraphics[width=\linewidth]{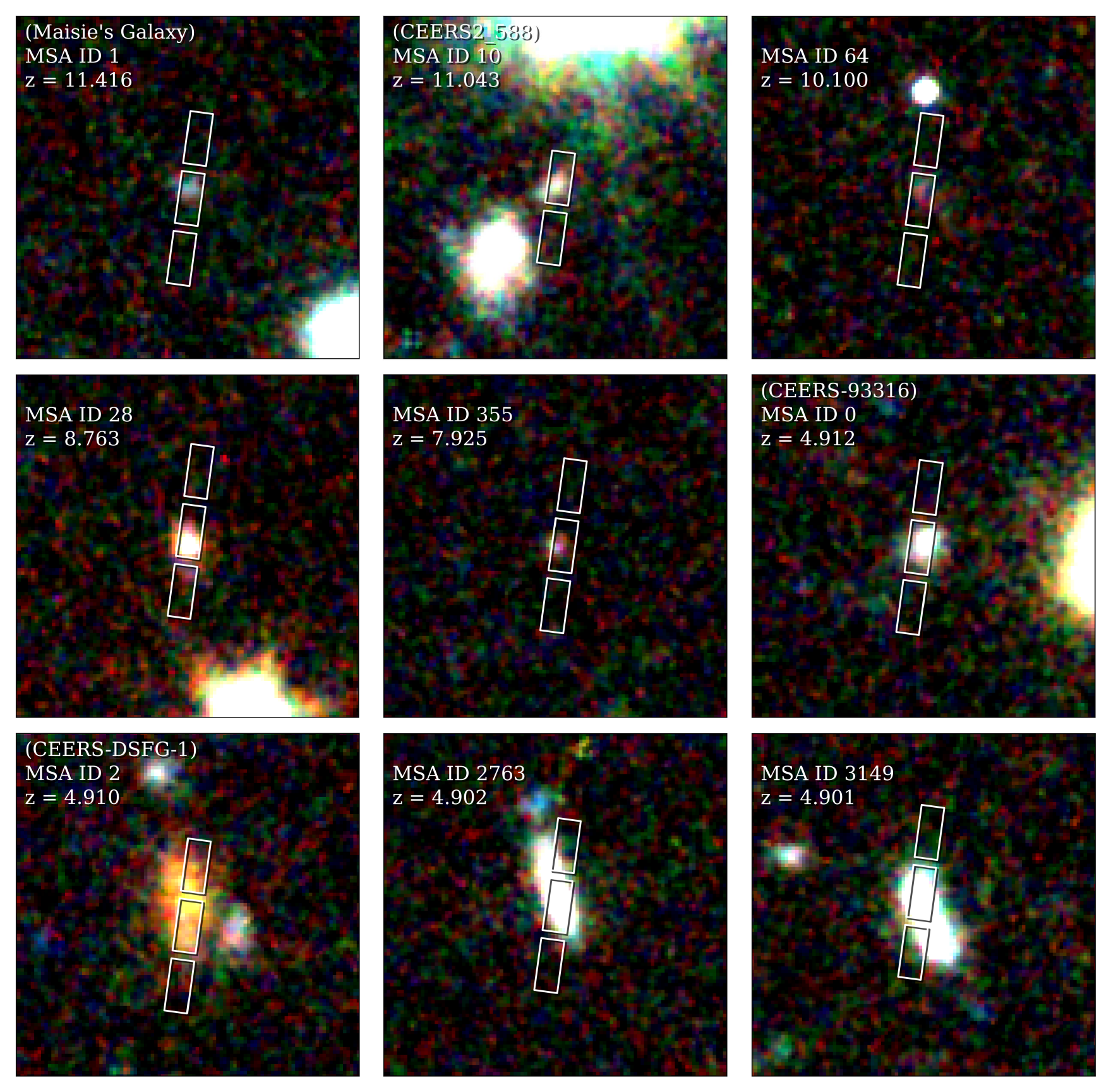}
    \caption{Montage of images of the galaxies targeted for MSA spectroscopy presented here. Each frame shows a $3^{\prime\prime} \times 3^{\prime\prime}$ stamp centered on the location of each galaxy, with ID labeled.  The red-green-blue image for each galaxy corresponds to the \textit{JWST}/NIRCam F444W, F356W, and F277W, respectively, rotated with North up and East to the left.  The rectangles show the approximate location of the \textit{JWST}/NIRSpec MSA slit positions.}
    \label{fig:montage}
\end{figure}

\begin{figure}
    \centering
    \includegraphics[width=\linewidth]{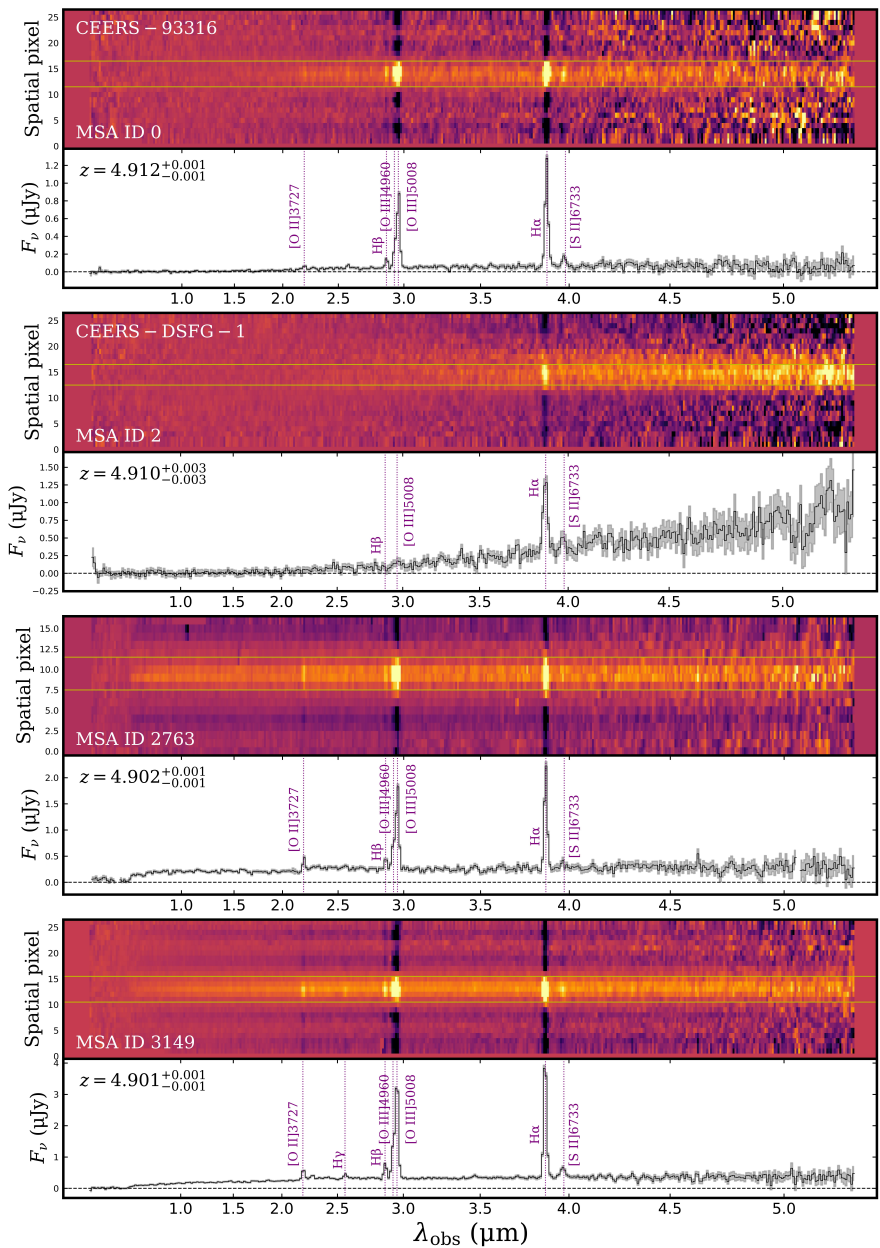}
    \caption{2D and 1D NIRSpec spectra of CEERS-93316, CEERS-DSFG-1 and two other galaxies at similar redshifts $z \approx 4.9$, highlighting a potential overdensity in the field at this redshift. The two horizontal yellow lines in the 2D spectra indicate the extraction aperture used to measure the 1D spectra. Dotted purple vertical lines indicate the location of several emission lines when present. The shaded grey area corresponds to 1$\sigma$ uncertainties in the 1D spectra and the zero flux level is marked by a dashed black horizontal line.}
    \label{fig:z4p9_spectra}
\end{figure}

For Maisie's Galaxy, CEERS2\_588, and MSA ID 64, we considered the alternative interpretation of a Balmer break at redshifts $z_\mathrm{BB} = 3.14$, 3.02 and 2.70, respectively.  Averaging the observed spectrum in $f_\lambda$ units over wavelength intervals $(1+z_\mathrm{BB}) \times$ 3145-3563~\AA\ and 3751-4198~\AA\ \citep{curtis-lake2023}, for Maisie's Galaxy we measure S/N of 0.0 and 9.7 below and above the break, and a $3\sigma$ limit on the flux ratio $> 2.6$, which is considerably larger than the maximum value of 1.8 that would be predicted from stellar population synthesis models \citep{curtis-lake2023}. This supports the higher redshift interpretation of the break resulting from Ly$\alpha$ opacity of the IGM. For CEERS2\_588 the spectrum is negative over the bluer window; the background may be slightly oversubtracted due to the presence of a large galaxy near the offset shutter in this two-shutter slitlet (see Extended Data Figure~\ref{fig:montage}).  The $3\sigma$ limits on the break amplitudes for CEERS2\_588 and 64 are $> 2.2$ and $>1.9$, respectively, also inconsistent with the Balmer break interpretation.

The derived spectroscopic redshifts are in broad agreement with the photometric redshift.  For Maisie's Galaxy, in particular, the photometric redshift has been modified since the galaxy was discovered as data processing improved.  The galaxy was originally thought to be at $z \sim$ 14, but the photometric redshift was revised to $z =$ 11.8$^{+0.2}_{-0.3}$ after improved astrometric registration recovered weak flux in F150W \cite{Finkelstein2022a}. Improved NIRCam zero-point flux calibration revised this further to $z =$ 11.5$^{+0.2}_{-0.6}$ \cite{Finkelstein2023}, while the latest photometric redshift from the photometry described in this paper (which is updated relative to earlier work \cite{Finkelstein2023} to better match {\it HST} and {\it JWST} photometry) suggests $z =$ 11.08$^{_+0.36}_{-0.39}$, consistent within the errors with the spectroscopic redshift $z = 11.416$.
For CEERS2\_588 the photometric ($z = 11.02^{+0.39}_{-0.27}$) and spectroscopic ($z=11.043$) redshifts also agree very well.

The Ly$\alpha$ break redshift we measure for CEERS2\_588 ($z = 11.43^{+0.01}_{-0.07}$) is significantly larger than the \oii\ emission line redshift.
A smaller offset with the same sense is found for Maisie's Galaxy. The spectral breaks in both galaxies are more gradual or ``rounded'' than is typical for  Ly$\alpha$ breaks observed in galaxies at lower redshifts, which may indicate the presence of a strong Ly$\alpha$ damping wing due to high column densities of neutral gas either intrinsic to (or immediately surrounding) the galaxies \citesupp{Heintz2023} or from a highly neutral IGM \citesupp{Umeda2023}.

CEERS-93316 and the millimeter-detected galaxy CEERS-DSFG-1 (MSA ID 2) both have redshifts $z = 4.91$.  We identify two more galaxies (MSA IDs 2763 and 3149) at nearly the same redshift in the NIRSpec observations (Extended Data Figure~\ref{fig:z4p9_spectra}, Extended Data Table~\hyperref[table:redshifts]{1}), supporting the hypothesis of a physical overdensity \citep{Naidu2022b}.  Moreover, the NOEMA-detected galaxy 1.5 arcsec from CEERS-93316 (Extended Data Figure~\ref{fig:noema}) has a very similar photometric redshift and may well also be associated, but still requires spectroscopy.

\subsection*{Emission lines}\label{sec:lines}

\begin{sidewaystable}
\begin{center}{\bf Extended Data Table 2: Emission line fluxes and limits}\end{center}
\begin{center}
\renewcommand{\arraystretch}{1} 
\begin{tabular}{lccccccccc}
\hline
\hline
MSA   & C \sc{iii}] & [O \sc{ii}] & [Ne \sc{iii}] & H$\gamma$ & H$\beta$ & [O \sc{iii}] & [O \sc{iii}] & H$\alpha$+[N \sc{ii}] & [S \sc{ii}]  \\ 
ID   & 1909 & 3727, 3730 & 3870 &  &  & 4960 & 5008 &  & 6718, 6733  \\ 
\hline 
   1  &  $ <42$  &  $28.5 \pm 6.7$  &  $ <21$  &  ---  &  ---  &  ---  &  ---  &  ---  &  ---  \\ 
  10  &  $ <66$  &  $45.2 \pm 5.9$  &  $ <21$  &  ---  &  ---  &  ---  &  ---  &  ---  &  ---  \\ 
  64  &  $ <46$  &  $ <17$  &  $ <17$  &  ---  &  ---  &  ---  &  ---  &  ---  &  ---  \\ 
  28  &  $79 \pm 15$  &  $56.6 \pm 5.0$  &  $55.7 \pm 7.7$  &  $64.6 \pm 7.8$  &  $103.2 \pm 6.4$  &  $204 \pm 28$  &  $601.7 \pm 8.8$  &  ---  &  ---  \\ 
 355  &  $ <63$  &  $ <13$  &  $ <13$  &  $7.0 \pm 2.2$  &  $25.6 \pm 4.5$  &  $56.6 \pm 1.3$  &  $166.3 \pm 4.0$  &  ---  &  ---  \\ 
   0  &  $ <189$  &  $91 \pm 17$  &  $ <52$  &  $54 \pm 12$  &  $101 \pm 10$  &  $256.9 \pm 4.0$  &  $755 \pm 12$  &  $538 \pm 11$  &  $60.0 \pm 8.3$  \\ 
   2  &  $ <768$  &  $ <227$  &  $ <210$  &  $ <163$  &  $ <129$  &  $ <126$  &  $ <125$  &  $667 \pm 30$  &  $ <92$  \\ 
2763  &  $ <417$  &  $396 \pm 34$  &  $ <141$  &  $ <106$  &  $198 \pm 24$  &  $485 \pm 11$  &  $1425 \pm 31$  &  $818 \pm 27$  &  $51 \pm 11$  \\ 
3149  &  $ <339$  &  $668 \pm 49$  &  $286 \pm 37$  &  $221 \pm 26$  &  $438 \pm 28$  &  $1023 \pm 13$  &  $3006 \pm 37$  &  $1674 \pm 27$  &  $181 \pm 10$  \\ 
\hline
\hline
\end{tabular}
\caption*{\textbf{Extended Data Table 2:} Integrated flux and limits for the main emission lines detected in the nine spectroscopically confirmed sources here presented, sorted by decreasing redshift as in Extended Data Table~\hyperref[table:redshifts]{1}.
All values are in units of 10$^{-20}$ erg s$^{-1}$ cm$^{-2}$ and all uncertainties listed are 68\% confidence intervals. Upper limits are presented at 3$\sigma$. The fluxes presented in this table have been measured after rescaling the spectra to the NIRCam photometry but note that small uncertainties due to imperfect photometric calibration and wavelength dependant slit losses could still be affecting this measurements and so they should be taken with caution.}
\label{table:emlines}
\end{center}
\end{sidewaystable}

The spectra of Maisie's Galaxy and CEERS2\_588 exhibit \oii\ emission lines with S/N = 4.9 and 6.4, respectively. While not highly significant, the emission lines are spatially well-aligned with the galaxies' continuum emission, and  careful inspection of subsets of data from the  nodded exposures show no indication that these features are data artifacts \citesupp{Harikane2023b}. 
At the expected wavelengths of the \neiii3870 line, CEERS2\_588 shows a feature with S/N = 2.0, and Maisie's Galaxy shows positive signal with still lower significance. Weak positive features are seen near the expected wavelengths for O\,{\sc iii}]1660 and \ciii1909, but these cannot be considered significant.

The emission lines in CEERS-93316 were fitted manually from the spectrum by first subtracting a local continuum and fitting Gaussian profiles to the visible emission lines.
In the case of the blended but partially-resolved \oiii4960,5008 lines, a double Gaussian function was fitted.
Fluxes and upper limits for the main emission lines present for the sources in this work are presented in Extended Data Table~\hyperref[table:emlines]{2}.
The Balmer lines (H$\alpha$ and H$\beta$) are corrected for underlying stellar absorption using the best-fitting stellar continuum from the Bagpipes fits (corrections $<1\%$; see below for a description of the Bagpipes fits).

H$\alpha$ is blended with \nii6550,6585 in the NIRSpec data.  We apply a correction \nii/H$\alpha$~$\approx$ 2\% based on an Oxygen abundance of $12+\log_{10}(\mathrm{O/H}) = 8.0$ estimated from the \oii, \oiii, H$\beta$ and \neiii\ lines.
The H$\alpha$ flux, corrected for \nii\ emission and for slit losses, is $(5.27 \pm 0.11) \times 10^{-18}$~erg~cm$^{-2}$~s$^{-1}$, 
and the Balmer decrement line flux ratio is $f(\mathrm{H}\alpha)/f(\mathrm{H}\beta) = 5.2 \pm 0.5$.
Assuming Case~B recombination at $T = 10^4$~K and Galactic extinction \citesupp{Cardelli1989} with $R_V = 3.1$, this corresponds to nebular extinction $\mathrm{E(B-V)}=0.60\pm0.10$ and a factor of $4.1^{+1.1}_{-0.9}$ attenuation at H$\alpha$. Assuming the nebular emission is produced by young star formation, this implies an extinction-corrected SFR(H$\alpha) = 28^{+7}_{-6}\ M_\odot$~yr$^{-1}$ \citepsupp{Kennicutt1998} for a Chabrier IMF \citepsupp{Chabrier2003}.

\subsection*{Spectral energy distribution modeling}\label{sec:SEDfit}

We use three spectral energy distribution (SED) fitting codes to derive physical properties for Maisie's Galaxy, CEERS2\_588, and CEERS-93316. SED fitting results are summarised in Extended Data Table~\hyperref[table:properties]{3}.

\begin{table}
\begin{center}{\bf Extended Data Table 3: Summary of galaxy physical properties}\end{center}
\begin{center}
\renewcommand{\arraystretch}{1.2} 
\begin{tabular}{lcccc}
\hline
\hline
  & MSA ID & Stellar Mass & SFR & Stellar $A_V$  \\
 &  & log$_{10}(M_{\star}/M_{\odot}$) 
& ($M_{\odot}$\,yr$^{-1}$) & (mag) 
 \\
\hline
Bagpipes \citesupp{carnall2018} 
 & 1 & $8.4^{+0.3}_{-0.4}$ & $3\pm2$ & $0.07^{+0.09}_{-0.05}$ \\
 & 10 & $8.5^{+0.3}_{-0.4}$ & $4\pm2$ & $0.10^{+0.11}_{-0.07}$ \\
 & 0 & $8.4\pm0.1$ & $3\pm1$ & $1.2\pm0.4$ \\
\hline
Cigale \citesupp{Burgarella2005, Boquien2019} 
 & 1 & 8.7 $\pm$ 0.3 & 8 $\pm$ 4 & 0.4 $\pm$ 0.4 \\
 & 10 & 8.8 $\pm$ 0.3 & 12 $\pm$ 5 & 0.4 $\pm$ 0.3 \\
 & 0 & 9.0 $\pm$ 0.2 & 18 $\pm$ 11 & 2.1 $\pm$ 0.6 \\
\hline
Synthesizer \citesupp{PerezGonzalez2003, PerezGonzalez2008}
 & 1 & 8.6 $\pm$ 0.3 & 2 $\pm$ 1 & 0.1 $\pm$ 0.1 \\
 & 10 &  8.7 $\pm$ 0.1  &  10 $\pm$ 4   &  0.2 $\pm$ 0.1 \\
 & 0 & 8.8 $\pm$ 0.1 & 60 $\pm$ 20 & 2.3 $\pm$ 0.2 \\

\hline
\hline
\end{tabular}
\caption*{\textbf{Extended Data Table 3:} Stellar population properties for Maisie's Galaxy (MSA ID 1, $z=11.416$), CEERS2\_588 (MSA ID 10, $z=11.043$) and CEERS-93316 (MSA ID 0, $z=4.912$) from the three different codes described in the Methods section. 
All uncertainties listed are 68\% confidence intervals.}
\label{table:properties}
\end{center}
\end{table}

For the two galaxies with $z > 11$ the three methods yield results for the stellar masses that are consistent within the uncertainties, and within a factor of about two for the best-fitting values. Dust attenuation values ($A_V$) are also consistent within the uncertainties, although Cigale finds a larger best-fit value with larger uncertainties.  The SFRs span larger ranges (factors of 3 to 4 between methods), likely due to different assumptions on allowable star formation histories (SFHs) and, for Cigale, the larger derived value for dust attenuation.

For CEERS-93316 the parameters derived by the three methods span larger ranges, most likely reflecting the complexity of this galaxy with its unusual combination of strong nebular line emission and large dust reddening. Bagpipes finds the lowest values for $M_\star$ and $A_V$, while Cigale and Synthesizer find higher values that are consistent within their uncertainties.  The large reddening implied for both the stars and the ionized gas is atypical for low mass ($M_\star < 10^9 M_\odot$) galaxies at high redshift \citepsupp{Pannella2015, Shapley2023}.
Bagpipes and Synthesizer find the lowest and highest values for the SFR, respectively, differing by a factor of 20, while Cigale's best fit SFR is intermediate between the other methods with a larger fractional uncertainty. 
The larger reddening and SFR values from Cigale and Synthesizer may arise because their fits are additionally constrained by the emission line EWs (and, with Synthesizer, the spectral continuum slope), whereas line emission influences Bagpipes only through its contribution to the NIRCam photometry.  Bagpipes also permits greater flexibility in the wavelength dependence of dust attenuation than was allowed for the other methods.

The SFR derived from the extinction-corrected H$\alpha$ emission from CEERS-93316 ($28^{+7}_{-6}\ M_\odot$~yr$^{-1}$) is on the high end of the range derived by Cigale, and below the range derived by Synthesizer. The evidence for strong reddening suggests the possibility that some star formation takes place in regions that are optically thick at UV-optical rest frame wavelengths, and hence that SFRs from SED fitting and/or H$\alpha$ emission may be underestimated.
The $3\sigma$ upper limit at 1.1~mm from NOEMA sets an upper limit to the dust-obscured SFR of $\sim 20$--200 $M_\odot$~yr$^{-1}$ assuming modified black body emission with dust temperatures in the range $T_{\rm d} = 40$--70~K and a typical spectral index of $\beta_{\rm d}=1.8$.  These limits are reasonably consistent with the SFR estimates derived from SED modeling and H$\alpha$, although the H$\alpha$ and Synthesizer SFRs would imply dust temperatures $T_d \gtrsim 41$~K and $\gtrsim 48$~K, respectively.

This analysis assumes that the light from CEERS-93316 is dominated by emission from stars and from gas ionized by star formation. Active galactic nuclei (AGN) can also be responsible for strong line emission, and indeed \textit{JWST} observations have  identified a significant population of reddened, broad line AGN at $z \sim 5$ \citep{kocevski23, Matthee2023}.  Most of these AGN are very compact or unresolved, whereas CEERS-93316 is spatially extended in NIRCam imaging, indicating that galaxy starlight is present, but this does not exclude a possible contribution from AGN emission as well. There is a tentative emission feature at the wavelength of \mgii2796,2804, a doublet frequently detected in AGN, but its significance is marginal (S/N = 1.8).

\subsubsection*{Bagpipes}

The Bagpipes \citesupp{carnall2018} fits make use of stellar population synthesis models \citesupp{Bruzual2003} assuming a Kroupa IMF \citesupp{Kroupa2001} with mass range 0.1--100~$M_\odot$, which results in masses and SFRs very similar to those from the Chabrier IMF \citesupp{Chabrier2003} used for the Cigale and Synthesizer fits. We consider stellar metallicities in the range $0.01 \leq Z/Z_\odot \leq 3.5$.  Gas metallicity is set equal to that of the stars.  Nebular emission lines and continuum are included in the model using results from the Cloudy photoionization code \citesupp{ferland2017}. Dust attenuation is included using a flexible-slope model, which also includes a Drude profile to model the 2175~\AA\ dust bump and an additional factor multiplying the attenuation applied to star-forming regions (which we take to include all stellar emission arising from stars younger than 10 Myr and all resulting nebular emission) \citesupp{salim2018}. The details of further Bagpipes model assumptions are identical to those used in previous work \citesupp{carnall2023, Donnan2023}. In particular, we use a constant SFR model, shown to produce stellar ages and stellar masses at the lower end of the plausible range for objects of interest. We apply a Gaussian prior to the redshift of each object based on the new spectroscopic data, with a standard deviation of 0.01 in each case. The median value for the Gaussian prior is set to $z=11.42$ for Maisie's Galaxy, $z=11.04$ for CEERS2\_588, and $z=4.91$ for CEERS-93316.  
For the Bagpipes fits, we use only galaxy photometry and spectroscopic redshifts, and do not include additional constraints from the spectra, such as emission line strengths.

\subsubsection*{Cigale}

We use the latest version of Cigale \citesupp{Burgarella2005,Boquien2019}, using stellar population synthesis models \citesupp{Stanway2018} assuming a Chabrier IMF \citesupp{Chabrier2003} with mass range 0.1--100~$M_\odot$. Stellar metallicities are allowed to vary from $0.0007 \leq Z/Z_{\odot} < 1$ for the $z > 11$ galaxies and $0.007 \leq Z/Z_{\odot} \leq 1$ for CEERS-93316.  Gas metallicites are allowed to vary independently from those of the stars, with a minimum value $10\times$ larger than the stellar minimum metallicity. We use updated emission line models computed from Cloudy \citepsupp{ferland2017}. A Calzetti extinction law \citesupp{Calzetti2000} is adopted for the dust attenuation of the stellar continuum. However, the nebular emission (continuum and lines) is attenuated with a screen model and a Small Magellanic Cloud (SMC) extinction curve \citepsupp[][]{Pei1992}. No 2175~\AA\ dust bump is added. 
For the $z > 11$ galaxies we that nebular extinction exceeds stellar extinction by a constant factor \citesupp{Calzetti2000}, while for CEERS-93316 we assume that young stars ($< 10$~Myr) and ionized gas are more heavily attenuated than older stars with a variable scaling factor \citesupp{Charlot2000}.
We select a \textit{``delayed + burst''} SFH. 
This form includes a delayed exponential term with $\mathrm{SFR} \propto t \times \exp(-t/\tau)$, with $\tau = 200$~Myr for the $z > 11$ galaxies and 500 and 1000 Myr for CEERS-93316.  On top of this, a short burst is allowed just prior to the time of observation. For the $z > 11$ galaxies the burst duration is 1~Myr and the burst mass fractions considered are 0\%, 0.5\%, and 10\%, whereas for CEERS-93316 we consider burst durations 5, 10 and 20~Myr and burst mass fractions 5\%, 7.5\%, 10\%, 12.5\% and 15\%.  For CEERS-93316, the best-fitting mass fraction is $10 \pm 3.5$\%;  the bursts are smaller for the $z > 11$ galaxies. The SFR reported here is the value averaged over 10~Myr prior to the time of observation.
The values for Gaussian priors on redshift were the same  as for the Bagpipes fitting.
Fitting with Cigale is constrained by the galaxy photometry and redshift.  Additionally, for CEERS-93316, we use constraints from the EWs measured from the NIRSpec spectrum for H$\alpha$, H$\beta$, and the two \oiii\ lines. 
The EWs from the best-fitting models match the observed values within 30\%.
The EW of the blended \oii\ doublet was included when fitting Maisie's Galaxy and CEERS2\_588.

\subsubsection*{Synthesizer}

With Synthesizer \citesupp{PerezGonzalez2003, PerezGonzalez2008}, we use stellar population synthesis models  \citesupp{Bruzual2003} assuming a Chabrier IMF \citesupp{Chabrier2003} with mass range 0.1--100~M$_\odot$.
We sample stellar metallicities in the range $0.005 \leq Z/Z_\odot \leq 1$.
Nebular continuum and hydrogen and helium emission lines are included in the modeling, assuming Case~B recombination, gas temperature $10^4$~K, density $10^2$~cm$^{-2}$, Lyman continuum photon escape fraction 15\%, Ly$\alpha$ forest transmission 10\% \citesupp{PerezGonzalez2008}. Conventional assumptions about the attenuation law and differential nebular-stellar dust effects are assumed \citesupp{Calzetti2000}, implying that the stellar continuum is less reddened than the nebular emission by a factor of $\sim$2. Attenuations are allowed to vary between $A_V=0$ and 1 mag for the $z>11$ galaxies, and up to 5~mag for CEERS-93316. 
We explore star formation histories described by a delayed exponential form (as also used for Cigale) with timescales ranging from 1~Myr  to 100~Myr in 0.1~dex steps. The SFR reported here is the value averaged over 100~Myr prior to the time of observation.
Redshifts are fixed to the values given in Extended Data Table~\hyperref[table:redshifts]{1}. 
Fitting with Synthesizer is constrained using both the galaxy photometry and the spectral fluxes, as well as the measured redshift.  
Additionally, for CEERS-93316, H$\alpha$ and H$\beta$ EWs measured from the spectra are included as a fitting constraint, but all models return lower values than those measured by a factor of $\sim 4$. For Maisie's Galaxy, a relatively strong degeneracy in the age is observed depending on the strength of the Balmer break, which is constrained by the spectroscopic data with relatively low S/N. Fitting spectral data points bluewards of the Balmer break yields younger mass-weighted ages and smaller stellar masses ($30 \pm 10$~Myr, $\log_{10}( M_{\star}/M_{\odot}) = 8.3 \pm 0.1$). Including NIRSpec spectroscopic data points redwards of the Balmer break increases the derived ages and masses ($120 \pm 30$~Myr, $\log_{10}( M_{\star}/M_{\odot}) = 8.8 \pm 0.1$).

\subsection*{Photometric redshift mimicry}\label{sec:zmimic}

The combination of very high EW nebular line emission, strongly reddened stellar continuum, and a very specific redshift value is responsible for CEERS-93316 mimicking the colors of a galaxy at $z \approx 16$. At $4.88 \lesssim z \lesssim 5.09$, H$\alpha$ falls in the ``triple overlap zone'' for all three of the reddest NIRCam bandpasses used in CEERS (F444W, F410M and F356W), while \oiii+H$\beta$ lies within the next bluest filter (F277W), as shown in Figure~\ref{fig:donnanspectrum}.  The emission lines boost the signal in those four bands to emulate a roughly flat $f_{\nu}$ color.  The highly reddened galaxy continuum fades at rest frame UV wavelengths 
($\lambda_\mathrm{obs} < 2.4\,\mu$m), without significant contribution from other emission lines, leading to a faint detection in F200W and no significant detections (at CEERS exposure times) in the bluer NIRCam filters or by \textit{HST}.  This leads to the mistaken impression of a very distant Ly$\alpha$ break ``dropout'' (Figure~\ref{fig:sed}).  Galaxies with very strong line emission are common at high redshift, but here the combination of strong line emission with a very red continuum is key to fooling the photometric redshift codes. A similar phenomenon was recognized for the submillimeter galaxy CEERS-DSFG-1 (MSA ID 2) \citep{zavala23}, which has possible (if not necessarily favored) photometric redshift solutions at $z_{\mathrm{phot}} > 12$, although the weak \oiii\ emission in that galaxy softens the effect compared to that for CEERS-93316. The other two galaxies with NIRSpec-confirmed redshifts $z \approx 4.9$ (MSA IDs 2763 and 3149) have bluer colors and are well-detected at NIRCam F115W and {\it HST} F814W, and therefore do not have viable photometric redshift solutions at $z_{\mathrm{phot}} > 6$. Galaxies like CEERS-93316 at other redshifts could produce similar photometric aliasing effects, depending on where their emission lines fall with respect to photometric bandpasses.  For example, at $6.71 < z < 6.99$ the \oiii5008 line would fall in the same NIRCam triple-overlap zone for F356W+F410M+F444W, mimicking the colors expected at $z \gtrsim 20$.  Observations through additional intermediate band filters could help to further distinguish low and high photometric redshift solutions for galaxies like these, but spectroscopy is the ultimate arbiter.

\subsection*{Data Availability}

The JWST NIRSpec data are available from the Mikulski Archive for Space Telescopes (MAST; \url{http://archive.stsci.edu}), under program ID 2750.  The CEERS JWST imaging data are available from MAST under program ID 1345. Reduced NIRCam data products from the CEERS team are available at \url{https://ceers.github.io}. The NOEMA data is part of program D22AC, which will be available at \url{http://vizier.cds.unistra.fr/viz-bin/VizieR-3?-source=B/iram/noema}.

\subsection*{Code Availability}
JWST NIRSpec data were reduced using the JWST Pipeline (Version 1.8.5, reference mapping 1069; \url{https://github.com/spacetelescope/jwst}). NIRSpec data inspection used the Mosviz visualization tool (\url{https://jdaviz.readthedocs.io/en/latest/mosviz/index.html} \citepsupp{JDADF2023}). Photometric redshifts were measured using {\sc eazy} \citesupp{brammer08} (\url{https://github.com/gbrammer/eazy-photoz/}).  Emission line profiles were fitted and line fluxes measured using {\sc LiMe} \citepsupp{Fernandez2023} (\url{https://lime-stable.readthedocs.io}). Three different codes were used for stellar populaation synthesis modeling of galaxy spectro-photometric data:  Synthesizer  \citesupp{PerezGonzalez2003,PerezGonzalez2008}; Bagpipes  \citesupp{carnall2018} (\url{https://bagpipes.readthedocs.io}); and Cigale \citesupp{Burgarella2005, Boquien2019} (\url{https://cigale.lam.fr}). For this paper, we use a non-public version of Cigale that fits photometric and spectroscopic data together.


\subsection*{Acknowledgments}

PAH, MED, SLF, JSK, and CJP acknowledge support from NASA through STScI ERS award JWST-ERS-1345.  SLF acknowledges support from the University of Texas at Austin.
PGP-G acknowledges support  from  Spanish  Ministerio  de  Ciencia e Innovaci\'on MCIN/AEI/10.13039/501100011033 through grant PGC2018-093499-B-I00. 
ACC acknowledges support from the Leverhulme Trust via a Leverhulme Early Career Fellowship.
CTD, DJM, RJM, and JSD acknowledge the support of the Science and Technology Facilities Council. FC acknowledges support from a UKRI Frontier Research Guarantee Grant [grant reference EP/X021025/1].
MHC acknowledges financial support from the State Research Agency (AEI\-MCINN) of the Spanish Ministry of Science and Innovation under the grants ``Galaxy Evolution with Artificial Intelligence" with reference PGC2018-100852-A-I00 and ``BASALT" with reference PID2021-126838NB-I00.
VF acknowledges support from ANID through FONDECYT Postdoc 2020 project 3200340. 
RA acknowledges support from ANID through FONDECYT Regular 1202007.

This work is based on observations with the NASA/ESA/CSA James Webb Space Telescope obtained from the Mikulski Archive for Space Telescopes at the STScI, which is operated by the Association of Universities for Research in Astronomy (AURA), Incorporated, under NASA contract NAS5-03127.

This work is partially based on observations carried out under project number D22AC with the IRAM NOEMA Interferometer. IRAM is supported by INSU/CNRS (France), MPG (Germany), and IGN (Spain).

\subsection*{Author contributions}
PAH led the \textit{JWST} NIRSpec DDT observing proposal, the observation design, the data reduction, and contributed extensively to text and figures.  MD, SLF, and JSK contributed extensively to spectroscopic target selection and prioritization, with extensive input from many authors including CTD, ACC, FC, JSD, SF, CP, PGP-G, DDK, DJM, RJM, and JAZ. Galaxy redshifts were measured primarily by PAH, VF, SLF, PGP-G, SF, IJ, and RLL.  RLL and VF measured emission line properties; FC and JRT analyzed implications from the emission lines.  DB, ACC, PGP-G and L-MS modeled and interpreted the spectral energy distributions of the galaxies. SF led the NOEMA observing program with contributions to planning from several coauthors including JAZ and CMC, and MK contributed to NOEMA execution and data analysis.  MD and SLF led the drafting of the  paper text with contributions to text, figures,  editing and formatting by PAH, SF, CP, PGP-G, HCF, MG, JAZ, TAH, KC, and other authors.  MBB led data reduction for the CEERS NIRCam observations with extensive contributions by AMK and HCF.  SHC and NP contributed to the NIRSpec proposal, and ROA, VB, MHC, DDK, RAL, and BJW (as well as many authors above) contributed to the discussion, analysis and interpretation.

\subsection*{Competing interests}
The authors declare that they have no competing financial interests. 

\subsection*{Additional information}
Correspondence and requests for materials should be addressed to PAH (email: \url{parrabalh@gmail.com}).


\begin{thebibliography}{10}
\expandafter\ifx\csname url\endcsname\relax
  \def\url#1{\texttt{#1}}\fi
\expandafter\ifx\csname urlprefix\endcsname\relax\def\urlprefix{URL }\fi
\providecommand{\bibinfo}[2]{#2}
\providecommand{\eprint}[2][]{\url{#2}}


\bibitem{schneider02}
\bibinfo{author}{{Schneider}, R.}, \bibinfo{author}{{Ferrara}, A.},
  \bibinfo{author}{{Natarajan}, P.} \& \bibinfo{author}{{Omukai}, K.}
\newblock \bibinfo{title}{{First Stars, Very Massive Black Holes, and Metals}}.
\newblock \emph{\bibinfo{journal}{\apj}} \textbf{\bibinfo{volume}{571}},
  \bibinfo{pages}{30--39} (\bibinfo{year}{2002}).

\bibitem{wise12}
\bibinfo{author}{{Wise}, J.~H.}, \bibinfo{author}{{Turk}, M.~J.},
  \bibinfo{author}{{Norman}, M.~L.} \& \bibinfo{author}{{Abel}, T.}
\newblock \bibinfo{title}{{The Birth of a Galaxy: Primordial Metal Enrichment
  and Stellar Populations}}.
\newblock \emph{\bibinfo{journal}{\apj}} \textbf{\bibinfo{volume}{745}},
  \bibinfo{pages}{50} (\bibinfo{year}{2012}).

\bibitem{jaacks19}
\bibinfo{author}{{Jaacks}, J.}, \bibinfo{author}{{Finkelstein}, S.~L.} \&
  \bibinfo{author}{{Bromm}, V.}
\newblock \bibinfo{title}{{Legacy of star formation in the pre-reionization
  universe}}.
\newblock \emph{\bibinfo{journal}{\mnras}} \textbf{\bibinfo{volume}{488}},
  \bibinfo{pages}{2202--2221} (\bibinfo{year}{2019}).

\bibitem{Castellano2022}
\bibinfo{author}{{Castellano}, M.} \emph{et~al.}
\newblock \bibinfo{title}{{Early Results from GLASS-JWST. III. Galaxy
  Candidates at z 9-15}}.
\newblock \emph{\bibinfo{journal}{\apjl}} \textbf{\bibinfo{volume}{938}},
  \bibinfo{pages}{L15} (\bibinfo{year}{2022}).


\bibitem{Finkelstein2022a}
\bibinfo{author}{{Finkelstein}, S.~L.} \emph{et~al.}
\newblock \bibinfo{title}{{A Long Time Ago in a Galaxy Far, Far Away: A
  Candidate z $\sim$ 12 Galaxy in Early JWST CEERS Imaging}}.
\newblock \emph{\bibinfo{journal}{\apjl}} \textbf{\bibinfo{volume}{940}},
  \bibinfo{pages}{L55} (\bibinfo{year}{2022}).

\bibitem{Donnan2023}
\bibinfo{author}{{Donnan}, C.~T.} \emph{et~al.}
\newblock \bibinfo{title}{{The evolution of the galaxy UV luminosity function
  at redshifts z $\sim$ 8 - 15 from deep JWST and ground-based near-infrared
  imaging}}.
\newblock \emph{\bibinfo{journal}{\mnras}} \textbf{\bibinfo{volume}{518}},
  \bibinfo{pages}{6011--6040} (\bibinfo{year}{2023}).

\bibitem{Harikane2023a}
\bibinfo{author}{{Harikane}, Y.} \emph{et~al.}
\newblock \bibinfo{title}{{A Comprehensive Study of Galaxies at z $\sim$ 9-16
  Found in the Early JWST Data: Ultraviolet Luminosity Functions and Cosmic
  Star Formation History at the Pre-reionization Epoch}}.
\newblock \emph{\bibinfo{journal}{\apjs}} \textbf{\bibinfo{volume}{265}},
  \bibinfo{pages}{5} (\bibinfo{year}{2023}).

\bibitem{Atek2023}
\bibinfo{author}{{Atek}, H.} \emph{et~al.}
\newblock \bibinfo{title}{{Revealing galaxy candidates out to z $\sim$ 16 with
  JWST observations of the lensing cluster SMACS0723}}.
\newblock \emph{\bibinfo{journal}{\mnras}} \textbf{\bibinfo{volume}{519}},
  \bibinfo{pages}{1201--1220} (\bibinfo{year}{2023}).

\bibitem{Finkelstein2023}
\bibinfo{author}{{Finkelstein}, S.~L.} \emph{et~al.}
\newblock \bibinfo{title}{{CEERS Key Paper. I. An Early Look into the First 500 Myr of Galaxy Formation with JWST}}.
\newblock \emph{\bibinfo{journal}{\apjl}} \textbf{\bibinfo{volume}{946}},
  \bibinfo{pages}{L13} (\bibinfo{year}{2023}).

\bibitem{boylankolchin23}
\bibinfo{author}{{Boylan-Kolchin}, M.}
\newblock \bibinfo{title}{{Stress testing {\ensuremath{\Lambda}}CDM with high-redshift galaxy candidates}}.
\newblock \emph{\bibinfo{journal}{Nat. Astron.}} \textbf{\bibinfo{volume}{7}},
  \bibinfo{pages}{731--735} (\bibinfo{year}{2023}).

\bibitem{Rieke2023}
\bibinfo{author}{{Rieke}, M.~J.} \emph{et~al.}
\newblock \bibinfo{title}{{Performance of NIRCam on JWST in Flight}}.
\newblock \emph{\bibinfo{journal}{\pasp}} \textbf{\bibinfo{volume}{135}},
  \bibinfo{pages}{028001} (\bibinfo{year}{2023}).

\bibitem{bagley23}
\bibinfo{author}{{Bagley}, M.~B.} \emph{et~al.}
\newblock \bibinfo{title}{{CEERS Epoch 1 NIRCam Imaging: Reduction Methods and Simulations Enabling Early JWST Science Results}}.
\newblock \emph{\bibinfo{journal}{\apjl}} \textbf{\bibinfo{volume}{946}},
  \bibinfo{pages}{L12} (\bibinfo{year}{2023}).

\bibitem{Robertson2023}
\bibinfo{author}{{Robertson}, B.~E.} \emph{et~al.}
\newblock \bibinfo{title}{{Identification and properties of intense star-forming galaxies at redshifts z $>$ 10}}.
\newblock \emph{\bibinfo{journal}{Nat. Astron.}} \textbf{\bibinfo{volume}{7}},
  \bibinfo{pages}{611--621} (\bibinfo{year}{2023}).

\bibitem{curtis-lake2023}
\bibinfo{author}{{Curtis-Lake}, E.} \emph{et~al.}
\newblock \bibinfo{title}{{Spectroscopic confirmation of four metal-poor galaxies at z = 10.3-13.2}}.
\newblock \emph{\bibinfo{journal}{Nat. Astron.}} \textbf{\bibinfo{volume}{7}},
  \bibinfo{pages}{622--632} (\bibinfo{year}{2023}).

\bibitem{Bouwens2023}
\bibinfo{author}{{Bouwens}, R.} \emph{et~al.}
\newblock \bibinfo{title}{{UV luminosity density results at z $>$ 8 from the first JWST/NIRCam fields: limitations of early data sets and the need for spectroscopy}}.
\newblock \emph{\bibinfo{journal}{\mnras}} \textbf{\bibinfo{volume}{523}},
  \bibinfo{pages}{1009--1035} (\bibinfo{year}{2023}).

\bibitem{Naidu2022b}
\bibinfo{author}{{Naidu}, R.~P.} \emph{et~al.}
\newblock \bibinfo{title}{{Schrodinger's Galaxy Candidate: Puzzlingly Luminous
  at $z\approx17$, or Dusty/Quenched at $z\approx5$?}}
\newblock \emph{\bibinfo{journal}{arXiv e-prints}}
  \bibinfo{pages}{arXiv:2208.02794} (\bibinfo{year}{2022}).

\bibitem{perezgonzalez23}
\bibinfo{author}{{P{\'e}rez-Gonz{\'a}lez}, P.~G.} \emph{et~al.}
\newblock \bibinfo{title}{{CEERS Key Paper. IV. A Triality in the Nature of HST-dark Galaxies}}.
\newblock \emph{\bibinfo{journal}{\apjl}} \textbf{\bibinfo{volume}{946}},
  \bibinfo{pages}{L16} (\bibinfo{year}{2023}).

\bibitem{zavala23}
\bibinfo{author}{{Zavala}, J.~A.} \emph{et~al.}
\newblock \bibinfo{title}{{Dusty Starbursts Masquerading as Ultra-high Redshift
  Galaxies in JWST CEERS Observations}}.
\newblock \emph{\bibinfo{journal}{\apjl}} \textbf{\bibinfo{volume}{943}},
  \bibinfo{pages}{L9} (\bibinfo{year}{2023}).

\bibitem{Jakobsen2022}
\bibinfo{author}{{Jakobsen}, P.} \emph{et~al.}
\newblock \bibinfo{title}{{The Near-Infrared Spectrograph (NIRSpec) on the
  James Webb Space Telescope. I. Overview of the instrument and its
  capabilities}}.
\newblock \emph{\bibinfo{journal}{\aap}} \textbf{\bibinfo{volume}{661}},
  \bibinfo{pages}{A80} (\bibinfo{year}{2022}).

\bibitem{ArrabalHaro2023}
\bibinfo{author}{{Arrabal Haro}, P.} \emph{et~al.}
\newblock \bibinfo{title}{{Spectroscopic Confirmation of CEERS NIRCam-selected Galaxies at $z \simeq 8-10$}}.
\newblock \emph{\bibinfo{journal}{\apjl}} \textbf{\bibinfo{volume}{951}},
  \bibinfo{pages}{L22} (\bibinfo{year}{2023}).
  
\bibitem{Harikane2023b}
\bibinfo{author}{{Harikane}, Y.} \emph{et~al.}
\newblock \bibinfo{title}{{Pure Spectroscopic Constraints on UV Luminosity Functions and Cosmic Star Formation History From 25 Galaxies at $z_\mathrm{spec}=8.61-13.20$ Confirmed with JWST/NIRSpec}}.
\newblock \emph{\bibinfo{journal}{arXiv e-prints}} 
  \bibinfo{pages}{arXiv:2304.06658} (\bibinfo{year}{2023}).

\bibitem{Fujimoto2023}
\bibinfo{author}{{Fujimoto}, S.} \emph{et~al.}
\newblock \bibinfo{title}{{CEERS Spectroscopic Confirmation of NIRCam-selected z {\ensuremath{\gtrsim}} 8 Galaxy Candidates with JWST/NIRSpec: Initial Characterization of Their Properties}}.
\newblock \emph{\bibinfo{journal}{\apjl}} \textbf{\bibinfo{volume}{949}},
  \bibinfo{pages}{L25} (\bibinfo{year}{2023}).

\bibitem{RobertsBorsani2023}
\bibinfo{author}{{Roberts-Borsani}, G.} \emph{et~al.}
\newblock \bibinfo{title}{{The nature of an ultra-faint galaxy in the cosmic dark ages seen with JWST}}.
\newblock \emph{\bibinfo{journal}{\nat}} \textbf{\bibinfo{volume}{618}},
  \bibinfo{pages}{480--483} (\bibinfo{year}{2023}).


\bibitem{Williams2023}
\bibinfo{author}{{Williams}, H.} \emph{et~al.}
\newblock \bibinfo{title}{{A magnified compact galaxy at redshift 9.51 with strong nebular emission lines}}.
\newblock \emph{\bibinfo{journal}{\sci}} \textbf{\bibinfo{volume}{380}},
  \bibinfo{pages}{416--420} (\bibinfo{year}{2023}).

\bibitem{Tacchella2023}
\bibinfo{author}{{Tacchella}, S.} \emph{et~al.}
\newblock \bibinfo{title}{{JADES Imaging of GN-z11: Revealing the Morphology
  and Environment of a Luminous Galaxy 430 Myr After the Big Bang}}.
\newblock \emph{\bibinfo{journal}{arXiv e-prints}}
  \bibinfo{pages}{arXiv:2302.07234} (\bibinfo{year}{2023}).

\bibitem{Barrufet2023}
\bibinfo{author}{{Barrufet}, L.} \emph{et~al.}
\newblock \bibinfo{title}{{Unveiling the nature of infrared bright, optically dark galaxies with early JWST data}}.
\newblock \emph{\bibinfo{journal}{\mnras}} \textbf{\bibinfo{volume}{522}},
  \bibinfo{pages}{449--456} (\bibinfo{year}{2023}).

\bibitem{kocevski23}
\bibinfo{author}{{Kocevski}, D.~D.} \emph{et~al.}
\newblock \bibinfo{title}{{Hidden Little Monsters: Spectroscopic Identification
  of Low-Mass, Broad-Line AGN at $z>5$ with CEERS}}.
\newblock \emph{\bibinfo{journal}{arXiv e-prints}}
  \bibinfo{pages}{arXiv:2302.00012} (\bibinfo{year}{2023}).


\bibitem{Matthee2023}
\bibinfo{author}{{Matthee}, J.} \emph{et~al.}
\newblock \bibinfo{title}{{Little Red Dots: an abundant population of faint AGN at $z\sim5$ revealed by the EIGER and FRESCO JWST surveys}}".
\newblock \emph{\bibinfo{journal}{arXiv e-prints}}
   \bibinfo{pages}{arXiv:2306.05448} (\bibinfo{year}{2023}).

\bibitem{bunker2023}
\bibinfo{author}{{Bunker}, A.~J.} \emph{et~al.}
\newblock \bibinfo{title}{{JADES NIRSpec Spectroscopy of GN-z11: Lyman-$\alpha$
  emission and possible enhanced nitrogen abundance in a $z=10.60$ luminous
  galaxy}}.
\newblock \emph{\bibinfo{journal}{arXiv e-prints}}
  \bibinfo{pages}{arXiv:2302.07256} (\bibinfo{year}{2023}).


\bibitem{Hsiao2023}
\bibinfo{author} {Hsiao, T.} \emph{et~al.}
\newblock \bibinfo{title}{JWST NIRSpec spectroscopy of the triply-lensed $z = 10.17$ galaxy MACS0647$-$JD}.
\newblock \emph{\bibinfo{journal}{arXiv e-prints}}
  \bibinfo{pages}{arXiv:2305.03042} (\bibinfo{year}{2023}).

\bibitem{Leung2023}
\bibinfo{author}{{Leung}, G. C. K.} \emph{et~al.}
\newblock \bibinfo{title}{{NGDEEP Epoch 1: The Faint-End of the Luminosity Function at $z \sim$ 9-12 from Ultra-Deep JWST Imaging}}
\newblock \emph{\bibinfo{journal}{arXiv e-prints}}
  \bibinfo{pages}{arXiv:2306.06244} (\bibinfo{year}{2023}).

\bibitem{Ferrara2023}
\bibinfo{author}{{Ferrara}, A.}, \bibinfo{author}{{Pallottini}, A.} \&
  \bibinfo{author}{{Dayal}, P.}
\newblock \bibinfo{title}{{On the stunning abundance of super-early, luminous galaxies revealed by JWST}}.
\newblock \emph{\bibinfo{journal}{\mnras}} \textbf{\bibinfo{volume}{522}},
  \bibinfo{pages}{3986--3991} (\bibinfo{year}{2023}).
  
\bibitem{Mason2023}
\bibinfo{author}{{Mason}, C.~A.}, \bibinfo{author}{{Trenti}, M.} \&
  \bibinfo{author}{{Treu}, T.}
\newblock \bibinfo{title}{{The brightest galaxies at cosmic dawn}}.
\newblock \emph{\bibinfo{journal}{\mnras}} \textbf{\bibinfo{volume}{521}},
  \bibinfo{pages}{497--503} (\bibinfo{year}{2023}).

\bibitem{bromm01}
\bibinfo{author}{{Bromm}, V.}, \bibinfo{author}{{Kudritzki}, R.~P.} \&
  \bibinfo{author}{{Loeb}, A.}
\newblock \bibinfo{title}{{Generic Spectrum and Ionization Efficiency of a
  Heavy Initial Mass Function for the First Stars}}.
\newblock \emph{\bibinfo{journal}{\apj}} \textbf{\bibinfo{volume}{552}},
  \bibinfo{pages}{464--472} (\bibinfo{year}{2001}).

\bibitem{sharda22}
\bibinfo{author}{{Sharda}, P.} \& \bibinfo{author}{{Krumholz}, M.~R.}
\newblock \bibinfo{title}{{When did the initial mass function become
  bottom-heavy?}}
\newblock \emph{\bibinfo{journal}{\mnras}} \textbf{\bibinfo{volume}{509}},
  \bibinfo{pages}{1959--1984} (\bibinfo{year}{2022}).

\bibitem{larson98}
\bibinfo{author}{{Larson}, R.~B.}
\newblock \bibinfo{title}{{Early star formation and the evolution of the
  stellar initial mass function in galaxies}}.
\newblock \emph{\bibinfo{journal}{\mnras}} \textbf{\bibinfo{volume}{301}},
  \bibinfo{pages}{569--581} (\bibinfo{year}{1998}).

\end{thebibliography}

\begin{thebibliography}{10}
\makeatletter
\addtocounter{\@listctr}{36}
\makeatother

\expandafter\ifx\csname url\endcsname\relax
  \def\url#1{\texttt{#1}}\fi
\expandafter\ifx\csname urlprefix\endcsname\relax\def\urlprefix{URL }\fi
\providecommand{\bibinfo}[2]{#2}
\providecommand{\eprint}[2][]{\url{#2}}


\bibitem{Planck2020}
\bibinfo{author}{Planck Collaboration}
\newblock \bibinfo{title}{Planck 2018 results. VI. Cosmological parameters}.
\newblock \emph{\bibinfo{journal}{\aap}} \textbf{\bibinfo{volume}{641}},
  \bibinfo{pages}{A6} (\bibinfo{year}{2020}).

\bibitem{Oke1983}
\bibinfo{author}{{Oke}, J.~B.} \& \bibinfo{author}{{Gunn}, J.~E.}
\newblock \bibinfo{title}{{Secondary standard stars for absolute spectrophotometry}}.
\newblock \emph{\bibinfo{journal}{\apj}} \textbf{\bibinfo{volume}{266}},
  \bibinfo{pages}{713--717} (\bibinfo{year}{1983}).
  
\bibitem{brammer08}
\bibinfo{author}{{Brammer}, G.~B.}, \bibinfo{author}{{van Dokkum}, P.~G.} \&
  \bibinfo{author}{{Coppi}, P.}
\newblock \bibinfo{title}{{EAZY: A Fast, Public Photometric Redshift Code}}.
\newblock \emph{\bibinfo{journal}{\apj}} \textbf{\bibinfo{volume}{686}},
  \bibinfo{pages}{1503--1513} (\bibinfo{year}{2008}).

\bibitem{larson22}
\bibinfo{author}{{Larson}, R.~L.} \emph{et~al.}
\newblock \bibinfo{title}{{Spectral Templates Optimal for Selecting Galaxies at
  z $>$ 8 with JWST}}.
\newblock \emph{\bibinfo{journal}{arXiv e-prints}}
  \bibinfo{pages}{arXiv:2211.10035} (\bibinfo{year}{2022}).

\bibitem{CASA2022}
\bibinfo{author}{{CASA Team}} \emph{et~al.}
\newblock \bibinfo{title}{{CASA, the Common Astronomy Software Applications for
  Radio Astronomy}}.
\newblock \emph{\bibinfo{journal}{\pasp}} \textbf{\bibinfo{volume}{134}},
  \bibinfo{pages}{114501} (\bibinfo{year}{2022}).

\bibitem{Ferruit2022}
\bibinfo{author}{{Ferruit}, P.} \emph{et~al.}
\newblock \bibinfo{title}{{The Near-Infrared Spectrograph (NIRSpec) on the
  James Webb Space Telescope. II. Multi-object spectroscopy (MOS)}}.
\newblock \emph{\bibinfo{journal}{\aap}} \textbf{\bibinfo{volume}{661}},
  \bibinfo{pages}{A81} (\bibinfo{year}{2022}).

\bibitem{JDADF2023}
\bibinfo{author}{{Developers}, J.} \emph{et~al.}
\newblock \bibinfo{title}{{Jdaviz}}.
\newblock \bibinfo{howpublished}{Zenodo} (\bibinfo{year}{2023}).

\bibitem{Storey_2004_O3}
\bibinfo{author}{Storey, P.~J.}, \bibinfo{author}{Sochi, T.} \&
  \bibinfo{author}{Badnell, N.~R.}
\newblock \bibinfo{title}{{Collision strengths for nebular [OIII] optical and
  infrared lines}}.
\newblock \emph{\bibinfo{journal}{\mnras}} \textbf{\bibinfo{volume}{441}}, \bibinfo{pages}{3028--3039}
  (\bibinfo{year}{2014}).

\bibitem{dijkstra14}
\bibinfo{author}{{Dijkstra}, M.}
\newblock \bibinfo{title}{{Ly{$\alpha$} Emitting Galaxies as a Probe of
  Reionisation}}.
\newblock \emph{\bibinfo{journal}{\pasa}} \textbf{\bibinfo{volume}{31}},
  \bibinfo{pages}{40} (\bibinfo{year}{2014}).


\bibitem{finkelstein19}
\bibinfo{author}{{Finkelstein}, S.~L.} \emph{et~al.}
\newblock \bibinfo{title}{{Conditions for Reionizing the Universe with a Low
  Galaxy Ionizing Photon Escape Fraction}}.
\newblock \emph{\bibinfo{journal}{\apj}} \textbf{\bibinfo{volume}{879}},
  \bibinfo{pages}{36} (\bibinfo{year}{2019}).


\bibitem{Heintz2023}
\bibinfo{author}{{Heintz}, K.} \emph{et~al.}
\newblock \bibinfo{title}{Extreme damped Lyman-$\alpha$ absorption in young star-forming galaxies at $z=9-11$}.
\newblock \emph{\bibinfo{journal}{arXiv e-prints}}
   \bibinfo{pages}{arXiv:2306.00647} (\bibinfo{year}{2023}).

\bibitem{Umeda2023}
\bibinfo{author}{{Umeda}, H.} \emph{et~al.}
\newblock \bibinfo{title}{{JWST Measurements of Neutral Hydrogen Fractions and Ionized Bubble Sizes at $z=7-12$ Obtained with Ly$\alpha$ Damping Wing Absorptions in 26 Bright Continuum Galaxies}}".
\newblock \emph{\bibinfo{journal}{arXiv e-prints}}
   \bibinfo{pages}{arXiv:2306.00487} (\bibinfo{year}{2023}).


\bibitem{Cardelli1989}
\bibinfo{author}{{Cardelli}, J.~A.}, \bibinfo{author}{{Clayton}, G.~C.} \& \bibinfo{author}{{Mathis}, J.~S.}
\newblock \bibinfo{title}{{The Relationship between Infrared, Optical, and Ultraviolet Extinction}}.
\newblock \emph{\bibinfo{journal}{\apj}} \textbf{\bibinfo{volume}{345}},
  \bibinfo{pages}{245} (\bibinfo{year}{1989}).

\bibitem{Kennicutt1998}
\bibinfo{author}{{Kennicutt}, R. C., Jr.}
\newblock \bibinfo{title}{{Star Formation in Galaxies Along the Hubble Sequence}}.
\newblock \emph{\bibinfo{journal}{\araa}} \textbf{\bibinfo{volume}{36}},
  \bibinfo{pages}{189-232} (\bibinfo{year}{1998}).

\bibitem{Chabrier2003}
\bibinfo{author}{{Chabrier}, G.}
\newblock \bibinfo{title}{{Galactic Stellar and Substellar Initial Mass
  Function}}.
\newblock \emph{\bibinfo{journal}{\pasp}} \textbf{\bibinfo{volume}{115}},
  \bibinfo{pages}{763--795} (\bibinfo{year}{2003}).

\bibitem{Pannella2015}
\bibinfo{author}{{Pannella}, M.} \emph{et~al.}
\newblock \bibinfo{title}{{GOODS-Herschel: Star Formation, Dust Attenuation, and the FIR-radio Correlation on the Main Sequence of Star-forming Galaxies up to $z \simeq 4$}}.
\newblock \emph{\bibinfo{journal}{\apj}} \textbf{\bibinfo{volume}{807}},
  \bibinfo{pages}{141} (\bibinfo{year}{2015}).

\bibitem{Shapley2023}
\bibinfo{author}{{Shapley}, A.~E.}, \bibinfo{author}{{Sanders}, R.~L.}, \bibinfo{author}{{Reddy}, N.~A.}, \bibinfo{author}{{Topping}, M.~W.} \& \bibinfo{author}{{Brammer}, G.~B.}
\newblock \bibinfo{title}{{JWST/NIRSpec Balmer-line Measurements of Star Formation and Dust Attenuation at $z \sim 3$-6}}".
\newblock \emph{\bibinfo{journal}{arXiv e-prints}}
   \bibinfo{pages}{arXiv:2301.03241} (\bibinfo{year}{2023}).

\bibitem{carnall2018}
\bibinfo{author}{{Carnall}, A.~C.}, \bibinfo{author}{{McLure}, R.~J.},
  \bibinfo{author}{{Dunlop}, J.~S.} \& \bibinfo{author}{{Dav{\'e}}, R.}
\newblock \bibinfo{title}{{Inferring the star formation histories of massive
  quiescent galaxies with BAGPIPES: evidence for multiple quenching
  mechanisms}}.
\newblock \emph{\bibinfo{journal}{\mnras}} \textbf{\bibinfo{volume}{480}},
  \bibinfo{pages}{4379--4401} (\bibinfo{year}{2018}).

\bibitem{Bruzual2003}
\bibinfo{author}{{Bruzual}, G.} \& \bibinfo{author}{{Charlot}, S.}
\newblock \bibinfo{title}{{Stellar population synthesis at the resolution of
  2003}}.
\newblock \emph{\bibinfo{journal}{\mnras}} \textbf{\bibinfo{volume}{344}},
  \bibinfo{pages}{1000--1028} (\bibinfo{year}{2003}).

\bibitem{Kroupa2001}
\bibinfo{author}{{Kroupa}, P.}
\newblock \bibinfo{title}{{On the variation of the initial mass function}}.
\newblock \emph{\bibinfo{journal}{\mnras}} \textbf{\bibinfo{volume}{322}},
  \bibinfo{pages}{231-246} (\bibinfo{year}{2001}).

\bibitem{ferland2017}
\bibinfo{author}{{Ferland}, G.~J.} \emph{et~al.}
\newblock \bibinfo{title}{{The 2017 Release Cloudy}}.
\newblock \emph{\bibinfo{journal}{\rmxaa}} \textbf{\bibinfo{volume}{53}},
  \bibinfo{pages}{385--438} (\bibinfo{year}{2017}).

\bibitem{salim2018}
\bibinfo{author}{{Salim}, S.}, \bibinfo{author}{{Boquien}, M.} \&
  \bibinfo{author}{{Lee}, J.~C.}
\newblock \bibinfo{title}{{Dust Attenuation Curves in the Local Universe:
  Demographics and New Laws for Star-forming Galaxies and High-redshift
  Analogs}}.
\newblock \emph{\bibinfo{journal}{\apj}} \textbf{\bibinfo{volume}{859}},
  \bibinfo{pages}{11} (\bibinfo{year}{2018}).

\bibitem{carnall2023}
\bibinfo{author}{{Carnall}, A.~C.} \emph{et~al.}
\newblock \bibinfo{title}{{A first look at the SMACS0723 JWST ERO:
  spectroscopic redshifts, stellar masses, and star-formation histories}}.
\newblock \emph{\bibinfo{journal}{\mnras}} \textbf{\bibinfo{volume}{518}},
  \bibinfo{pages}{L45--L50} (\bibinfo{year}{2023}).


\bibitem{Boquien2019}
\bibinfo{author}{{Boquien}, M.} \emph{et~al.}
\newblock \bibinfo{title}{{CIGALE: a python Code Investigating GALaxy
  Emission}}.
\newblock \emph{\bibinfo{journal}{\aap}} \textbf{\bibinfo{volume}{622}},
  \bibinfo{pages}{A103} (\bibinfo{year}{2019}).

\bibitem{Burgarella2005}
\bibinfo{author}{{Burgarella}, D.}, \bibinfo{author}{{Buat}, V.} \&
  \bibinfo{author}{{Iglesias-P{\'a}ramo}, J.}
\newblock \bibinfo{title}{{Star formation and dust attenuation properties in
  galaxies from a statistical ultraviolet-to-far-infrared analysis}}.
\newblock \emph{\bibinfo{journal}{\mnras}} \textbf{\bibinfo{volume}{360}},
  \bibinfo{pages}{1413--1425} (\bibinfo{year}{2005}).

\bibitem{Stanway2018}
\bibinfo{author}{{Stanway}, E.~R.} \& \bibinfo{author}{{Eldridge}, J.~J.}
\newblock \bibinfo{title}{{Re-evaluating old stellar populations}}.
\newblock \emph{\bibinfo{journal}{\mnras}} \textbf{\bibinfo{volume}{479}},
  \bibinfo{pages}{75-93} (\bibinfo{year}{2018}).

\bibitem{Calzetti2000}
\bibinfo{author}{{Calzetti}, D.} \emph{et~al.}
\newblock \bibinfo{title}{{The Dust Content and Opacity of Actively
  Star-forming Galaxies}}.
\newblock \emph{\bibinfo{journal}{\apj}} \textbf{\bibinfo{volume}{533}},
  \bibinfo{pages}{682--695} (\bibinfo{year}{2000}).

\bibitem{Pei1992}
\bibinfo{author}{{Pei}, Y.~C.}
\newblock \bibinfo{title}{{Interstellar Dust from the Milky Way to the
  Magellanic Clouds}}.
\newblock \emph{\bibinfo{journal}{\apj}} \textbf{\bibinfo{volume}{395}},
  \bibinfo{pages}{130} (\bibinfo{year}{1992}).

\bibitem{Charlot2000}
\bibinfo{author}{{Charlot}, S.} \& \bibinfo{author}{{Fall}, S.~M.}
\newblock \bibinfo{title}{{A Simple Model for the Absorption of Starlight by Dust in Galaxies}}.
\newblock \emph{\bibinfo{journal}{\apj}} \textbf{\bibinfo{volume}{539}},
  \bibinfo{pages}{718-731} (\bibinfo{year}{2000}).

\bibitem{PerezGonzalez2003}
\bibinfo{author}{{P{\'e}rez-Gonz{\'a}lez}, P.~G.} \emph{et~al.}
\newblock \bibinfo{title}{{Stellar populations in local star-forming galaxies -
  I. Data and modeling procedure}}.
\newblock \emph{\bibinfo{journal}{\mnras}} \textbf{\bibinfo{volume}{338}},
  \bibinfo{pages}{508--524} (\bibinfo{year}{2003}).

\bibitem{PerezGonzalez2008}
\bibinfo{author}{{P{\'e}rez-Gonz{\'a}lez}, P.~G.} \emph{et~al.}
\newblock \bibinfo{title}{{The Stellar Mass Assembly of Galaxies from z = 0 to
  z = 4: Analysis of a Sample Selected in the Rest-Frame Near-Infrared with
  Spitzer}}.
\newblock \emph{\bibinfo{journal}{\apj}} \textbf{\bibinfo{volume}{675}},
  \bibinfo{pages}{234--261} (\bibinfo{year}{2008}).
  

\bibitem{Fernandez2023}
\bibinfo{author}{Fernández, V.}
\newblock \bibinfo{title}{{The resolved chemical composition of the starburst
  dwarf galaxy CGCG007-025: direct method versus photoionization model
  fitting}}.
\newblock \emph{\bibinfo{journal}{\mnras}} \textbf{\bibinfo{volume}{520}}, \bibinfo{pages}{3576--3590}
  (\bibinfo{year}{2023}).


\end{thebibliography}
\end{document}